\documentclass[12pt]{article}
\usepackage{epsfig,amsmath,amssymb,mathrsfs}

  {\end{list}}%

\makeatletter
\renewcommand{\section}{\setcounter{equation}{0}\@startsection
 {section}%
 {1}%
 {0pt}%
 {-1\baselineskip}%
 {0.4\baselineskip}%
 {\bfseries\large}}%
\renewcommand{\subsection}{\@startsection
 {subsection}%
 {2}%
 {0pt}%
 {-0.75\baselineskip}%
 {0.2\baselineskip}%
 {\bfseries}}%
\renewcommand{\subsubsection}{\@startsection
 {subsubsection}%
 {3}%
 {0pt}%
 {-0.5\baselineskip}%
 {0.1\baselineskip}%
 {\sc}}%
\makeatother

\DeclareMathAlphabet{\mathpzc}{OT1}{pzc}{m}{it}

\def\A{{\rm A}}
\def\B{{\rm B}}

\def\g5{\gamma_{5}}

\def\idpq{\int\!\! \frac{d^4\!p}{(2\pi)^4}\frac{d^4\!q}{(2\pi)^4}}

\def\idx{\int\!\! d^4\!x}

\def\unit{{\rm 1\! I}}

\newcommand{\bea}{\begin{eqnarray}}
\newcommand{\eea}{\end{eqnarray}}
\newcommand{\beann}{\begin{eqnarray*}}
\newcommand{\eeann}{\end{eqnarray*}}
\newcommand{\ba}{\begin{array}}
\newcommand{\ea}{\end{array}}

 \def\g {\gamma}
\begin{document}
 \begin{titlepage}
\rightline{FTI/UCM 37-2015}
\vglue 33pt

\begin{center}

{\Large \bf The hybrid Seiberg-Witten map, its $\theta$-exact expansion and the antifield formalism.}\\
\vskip 1.0true cm
{\rm C. P. Mart\'{\i}n}\footnote{E-mail: carmelop@fis.ucm.es} and David G. Navarro\footnote{E-mail: dgnavarro@ucm.es}
\vskip 0.1 true cm
{\it Departamento de F\'{\i}sica Te\'orica I,
Facultad de Ciencias F\'{\i}sicas\\
Universidad Complutense de Madrid,
 28040 Madrid, Spain}\\
\vskip 0.85 true cm
{\leftskip=50pt \rightskip=50pt \noindent
We deduce an evolution equation for an arbitrary hybrid Seiberg-Witten map for compact gauge groups by using the antifield formalism. We show how this evolution equation can be
used to obtain the hybrid Seiberg-Witten map as an expansion, which is $\theta$-exact, in the number of ordinary fields. We compute explicitly this expansion up to order three in the number of ordinary gauge fields and then particularize it to case of the  Higgs of the noncommutative Standard Model.
\par}
\end{center}

\vspace{9pt}
\noindent{\em PACS:} 11.10.Nx; 12.10.-g, 11.15.-q; \\
{\em Keywords:}  Noncommutative gauge theories, Seiberg-Witten map, $\theta$-exact.
\vfill
\end{titlepage}

\section{Introduction}
The Seiberg-Witten map was introduced in ref.~\cite{Seiberg:1999vs} to account for the fact that at the classical level the same underlying field theory can be defined by using either noncommutative gauge fields or ordinary gauge fields. Indeed, when noncommutative gauge fields are used to define the theory, the classical action is a polynomial with regard to the $\star$-product of the noncommutative gauge fields and their derivatives and  it is, the classical action, invariant under noncommutative $U(n)$ gauge transformations. However, this action turns out to contain an infinity of terms with ever increasing powers of the noncommutativity parameters, when ordinary gauge fields are employed to define it. The action in question is invariant under ordinary $U(n)$gauge transformations, when expressed in terms of the ordinary fields.

Strictly speaking, before the formalism proposed in Refs.~\cite{Madore:2000en,Jurco:2000ja,Jurco:2001rq} came about, the Standard Model of particle interactions had no counterpart  on noncommutative space-time --see, though, ref.~\cite{Khoze:2004zc} for a close relative of the Standard Model. The formalism in question is called the enveloping-algebra formalism  because the noncommutative gauge fields take values in the enveloping algebra of the Lie algebra of the corresponding ordinary gauge theory. In the enveloping-algebra formalism the noncommutative gauge fields are defined in terms of the ordinary gauge
fields by using a Seiberg-Witten map, and thus the ordinary infinitesimal gauge orbits are mapped into infinitesimal noncommutative ones. Noncommutative matter fields are defined in terms of the ordinary gauge fields and matter fields by using the appropriate Seiberg-Witten
map.  By employing the enveloping-algebra formalism the noncommutative counterpart of the Standard Model of particle interactions was finally formulated in ref.~\cite{Calmet:2001na}.
Some phenomenological consequences that arise when the Standard Model is formulated on noncommutative space-time have been analyzed in Refs.~\cite{Melic:2005su, Alboteanu:2006hh, Buric:2007qx, Tamarit:2008vy,Trampetic:2009vy, Haghighat:2010up, Wang:2011ei, YaserAyazi:2012ni, Aghababaei:2013dia, Ghasemkhani:2014eva, Haghighat:2014nra, Fresneda:2015zya}. The general construction of noncommutative GUTs was discussed in ref.~\cite{Aschieri:2002mc} and concrete examples were given in Refs.~\cite{Martin:2013gma, Martin:2013lba}. The Seiberg-Witten map has also been instrumental  in the formulation
of noncommutative gravity theories: see, for instance, Refs.~\cite{Calmet:2006iz, Marculescu:2008gw,Aschieri:2011ng, Aschieri:2012in, Dimitrijevic:2014iwa, Aschieri:2014xka}.

If the Seiberg-Witten map is computed by expanding the noncommutative fields in powers of the noncommutativity parameters and only a finite number of those terms are considered in the computations, one misses the UV/IR mixing effects that are a key feature~\cite{Minwalla:1999px, Hayakawa:1999yt} of noncommutative gauge theories when formulated in terms of the noncommutative fields. It was shown in ref.~\cite{Schupp:2008fs} that if the Seiberg-Witten map  is defined as an expansion in  powers of the coupling constant, or as an expansion in the number of ordinary fields, the UV/IR mixing effects do occur also when the noncommutative theory is expressed in terms of the ordinary fields; provided no expansion in powers of the noncommutativity parameters is carried out. This Seiberg-Witten map, where there is  no expansion in the noncommutativity parameters, is referred to as the $\theta$-exact Seiberg-Witten map. Several very interesting studies of the properties and phenomenological implications of the noncommutative field theories defined by means of the $\theta$-exact Seiberg-Witten map have been carried out so far --see Refs.~\cite{Horvat:2011iv, Horvat:2011qn,Horvat:2011qg,Horvat:2012vn,Horvat:2013rga, Trampetic:2015zma}, but much work is still waiting to be done.

The computation of the $\theta$-exact Seiberg-Witten map by brute force --ie, by coming up with an ansatz that solves the Seiberg-Witten map equation-- for nonabelian gauge groups is a daunting task due to the highly involved non polynomial dependence of the map on the momenta. In ref.~\cite{Martin:2012aw},
it was put forward a recursive  method to construct a $\theta$-exact Seiberg-Witten map for arbitrary gauge groups. The method in question produces a solution to the ``evolution" Seiberg-Witten map equation, an equation which was obtained in Refs.~\cite{Barnich:2001mc, Barnich:2002pb, Barnich:2002tz, Barnich:2003wq} by using the antifield formalism techniques --see Refs.~\cite{Gomis:2000sp, Brace:2001fj,Picariello:2001mu}, for alternative cohomogical approaches and also ref.4\cite{Ulker:2007fm}. However, there is an important type of Seiberg-Witten map which was not considered in ref.~\cite{Martin:2012aw} and whose ``evolution" equation has not been derived neither in Refs.~\cite{Barnich:2001mc, Barnich:2002pb, Barnich:2003wq} nor elsewhere. This type of Seiberg-Witten map is called the hybrid Seiberg-Witten map --see ref.~\cite{Schupp:2001we}-- and it is needed when we have noncommutative matter fields on which some noncommutative gauge transformations act from the left and others act from the right. The hybrid Seiberg-Witten map is a must when one wants to analyze, using ordinary fields, noncommutative theories with noncommutative fields which transforms under the fundamental representation of the Lie algebra of $U(n_L)$ on the left and under the fundamental representation of Lie algebra of $U(n_R)$ on the right. Actually,  the concept of hybrid Seiberg-Witten map was introduced in  ref.~\cite{Calmet:2001na} to construct the noncommutative Yukawa terms of the noncommutative Standard Model. Generally speaking, a noncommutative Yukawa term demands the existence of a hybrid Seiberg-Witten map for it to be expressible in terms of ordinary fields~\cite{Aschieri:2002mc, Martin:2010ng}.

The purpose of this paper is threefold. First, to obtain, by using the antifield techniques of Refs.~\cite{Barnich:2001mc, Barnich:2002pb, Barnich:2002tz, Barnich:2003wq}, an ``evolution" equation for a general hybrid Seiberg-Witten map. The reason why we shall use the anti-field formalism, and not a
 more direct method as in ref.~\cite{Seiberg:1999vs}, is that we want to fill a non-negligible gap that exists in the current literature. Indeed, we want to show that noncommutative gauge theories where there is a hybrid Seiberg-Witten map --the noncommutative Standard- Model, in particular-- also fall in the category of consistent deformations of gauge theories as defined in Ref.~\cite{Barnich:2001mc} by using the fruitful anti-field formalism and, hence, that the hybrid Seiberg-Witten map corresponds to an anticanonical transformation. This approach -the consistent deformation one-- to the formulation of noncommutative gauge theories has proved to be very illuminating and played a chief role~\cite{Brandt:2003fx} in the proof of the triviality of the $\theta$-dependent contributions to the noncommutative gauge anomaly expanded in powers of $\theta$. Second, to show that it can be solved recursively in Fourier space by carrying out a formal expansion of the noncommutative fields in terms of the number of ordinary gauge fields. Thus, no expansion in the noncommutativity parameters is introduced. Third, to work out the $\theta$-exact expression for a general hybrid Seiberg-Witten map up to order three in the number of ordinary gauge fields and particularize them to the noncommutative Higgs fields that occur in the noncommutative Standard Model of ref.~\cite{Calmet:2001na}. It should be stressed that defining the Seiberg-Witten map as a formal expansion in the number of ordinary gauge fields is quite in keeping with a formulation of the corresponding quantum field theory in terms of Feynman diagrams.

The layout of this paper is as follows. In Section 1, we derive by using the antifield formalism an ``evolution" equation which defines a general hybrid Seiberg-Witten map. In   section 2, we show how solve recursively the hybrid Seiberg-Witten map ``evolution" equation by expanding in the number of gauge
fields in Fourier space. The resulting general hybrid Seiberg-Witten map is worked out explicitly up to order three in the number of gauge fields. Then,  the general formulas are particularized to the Standard Model Higgs case and a $\theta$-exact expression is obtained for the type of Yukawa terms
that occur in the noncommutative Standard Model. Several appendices are included, which contain lengthy expressions not given in  the main sections of the paper.

\section{The hybrid Seiberg-Witten map and the antifield formalism}

Let $L_a$ and $R_a$ denote the generators, in arbitrary faithful finite dimensional matrix unitary representations, of  compact Lie groups $G_L$ and $G_R$, respectively. $L_a$ and $R_a$ will be hermitian matrices of dimension $n_L$ and $n_R$, respectively. Let $a_\mu(x)=a^a_\mu(x) L_a$ and $b_\mu(x)=b^a_\mu(x) R_a$ be ordinary gauge fields whose BRST transformations read
\begin{equation*}
sa_\mu=\partial_\mu\lambda+i[a_\mu,\lambda],\quad s\,b_\mu=\partial_\mu\omega+i[b_\mu,\omega],
\end{equation*}
where $\lambda(x)=\lambda^a(x) L_a$ and $\omega(x)=\omega^a(x) R_a$ denote the corresponding ordinary ghost fields. Let $\phi(x)$ denote an ordinary scalar field  which transforms as follows
\begin{equation*}
s\phi=-i\,\lambda\phi\,+\,i\,\phi\omega,
\end{equation*}
under the BRST transformations that $G_L$ --acting from the left--  and $G_R$ --acting from the right-- give  rise to.

Notice that $\phi(x)$ is valued in the space of $n_L\times n_R$ complex matrices; where $n_L$ and $n_R$ are the dimensions of the matrices which represent $L_a$ and $R_a$, respectively. Let us point out that it will become clear that the  Seiberg-Witten map ``evolution" equations presented below remain valid when $\phi(x)$ is a fermion field, but that we shall take $\phi(x)$ to be a scalar to avoid the proliferation of indices.

Let the Moyal product, $\star_{h}$, of two functions, $f_1$ and $f_2$, be defined as follows:
\begin{equation*}
(f_1\star_{h}f_2)(x)=\idpq\;{\tilde f}_1(p){\tilde f}_2(q)\; e^{-i\frac{h}{2}(p\wedge q)}\;e^{-i(p+q)x},
\end{equation*}
where $p\wedge q =\theta^{ij}\,p_{i}q_{j}$. ${\tilde f}_1$ and ${\tilde f}_2$ are the Fourier transforms of $f_1$ and $f_2$, respectively.

In the enveloping-algebra formalism~\cite{Jurco:2001rq}, to the  ordinary gauge fields $a_\mu$ and its ghost field $\lambda$, one associates a noncommutative gauge field, $A_\mu$,  and a noncommutative ghost field $\Lambda$, respectively. $A_\mu=A_\mu[a_\rho,\theta]$ and
$\Lambda=\Lambda[a_\mu,\lambda;\theta]$ are  functions of $a_\mu$, $\lambda$ and $\theta^{ij}$, such that they are a solution to the Seiberg-Witten map equations
\begin{equation}
\begin{array}{l}
{s_{NC}A_\mu[a_\rho;\theta] = s A_\mu[a_\rho;\theta],\quad s_{NC}\Lambda[a_\rho,\lambda;\theta]=s\Lambda[a_\rho,\lambda;\theta],}\\[4pt]
{A_\mu[a_\rho,\theta=0] = a_\mu,\quad \Lambda[a_\rho,\lambda;\theta=0]=\lambda.}
\end{array}
\label{Asweq}
\end{equation}
Above, the symbol $s_{NC}$ denotes the noncommutative BRST operator, which, by definition, acts on $A_\mu$ and $\Lambda$ as follows:
\begin{equation}
s_{NC}\,A_\mu=\partial_\mu\Lambda+i[A_\mu,\Lambda]_{\star_h},\quad s_{NC}\,\Lambda=-i\,\Lambda\star_h\Lambda.
\label{ABRST}
\end{equation}

Analogously, one associates to the ordinary gauge field $b_\mu$ and its ghost field $\omega$, a noncommutative field, $B_\mu=B_\mu[b_\rho,\theta]$, and a noncommutative ghost field,   $\Omega=\Omega[b_\rho,\omega;\theta]$. $B_\mu[b_\rho,\theta]$ and
$\Omega[b_\rho,\omega;\theta]$ are a solution to
\begin{equation}
\begin{array}{l}
{s_{NC}B_\mu[b_\rho;\theta] = s B_\mu[b_\rho;\theta],\quad s_{NC}\Omega[b_\rho,\omega;\theta]=s\omega[b_\rho,\omega;\theta],}\\[4pt]
{B_\mu[b_\rho,\theta=0] = b_\mu,\quad \omega[b_\rho,\omega;\theta=0]=\omega.}
\end{array}
\label{Bsweq}
\end{equation}
The action on $s_{NC}$ on $B_\mu$ and $\Omega$ is defined thus
\begin{equation}
s_{NC}\,B_\mu=\partial_\mu\Omega+i[B_\mu,\Omega]_{\star_h},\quad s_{NC}\,\Omega=-i\,\Omega\star_h\Omega.
\label{BBRST}
\end{equation}

Following Ref.~\cite{Schupp:2001we}, we shall associate a noncommutative field, $\Phi$, to the ordinary field $\phi$. We shall assume that $\Phi=\Phi[\phi,a_\rho,b_\rho;\theta]$ is given by formal power series of the ordinary fields $\phi$, $a_\mu$ and $b_\mu$ such that it satisfies the following equations
\begin{equation}
s_{NC}\Phi[\phi,a_\rho,b_\rho;\theta]=s\Phi[\phi,a_\rho,b_\rho;\theta],\quad \Phi[\phi,a_\rho,b_\rho;\theta=0]=\phi,
\label{hybrideq}
\end{equation}
where
\begin{equation}
s_{NC}\,\Phi\,=\,-i\,\Lambda\star_{h}\Phi\,+\,i\,\Phi\star_{h}\Omega,
\label{HBRST}
\end{equation}
with $\Lambda$ and $\Omega$ being the noncommutative ghost fields defined by (\ref{Asweq}) and (\ref{Bsweq}), respectively. A $\Phi=\Phi[\phi,a_\rho,b_\rho;\theta]$ that solves (\ref{hybrideq}) is called a hybrid Seiberg-Witten map. This map defines the
noncommutative  field $\Phi$ in terms of the ordinary field $\phi$, $a_\mu$ and $b_\mu$ in such a way that maps the ordinary infinitesimal gauge orbit of $\phi$ into the noncommutative infinitesimal gauge orbit of $\Phi$.

To construct real actions one also needs the hermitian conjugate of $\Phi$ and $\phi$, which we shall denote by $\bar{\Phi}$ and
$\bar{\phi}$, respectively.  As for the BRST transformations of $\bar{\Phi}$ and $\bar{\phi}$, we shall demand that
\begin{equation*}
\begin{array}{l}
{s_{NC}\,\bar{\Phi}\,=\,i\,\bar{\Phi}\star_{h}\Lambda\,-\,i\,\Omega\star_h\bar{\Phi},\quad
s\,\bar{\phi}\,=\,i\,\bar{\phi}\lambda\,-\,i\,\omega\bar{\phi},}\\[8pt]
{s_{NC}\bar{\Phi}[\bar{\phi},a_\rho,b_\rho;\theta]=s\bar{\Phi}[\bar{\phi},a_\rho,b_\rho;\theta],\quad \bar{\Phi}[\bar{\phi},a_\rho,b_\rho;\theta=0]=\bar{\phi},}
\end{array}
\end{equation*}
do hold.

The purpose of the current Section  is to show that a solution to the hybrid Seiberg-Witten map equations in (\ref{hybrideq}) --ie, a Seiberg-Witten map-- can be found by solving the following ``evolution" problem:
\begin{equation}
\begin{array}{l}
{{\displaystyle \frac{d\Phi}{dh}}=\frac{1}{2}\,\theta^{ij}\,A_i\star_{h}\partial_j\Phi+\frac{i}{4}\,\theta^{ij}\,A_i\star_{h} A_j\star_{h}\Phi}\\[8pt]
{\phantom{{\displaystyle \frac{d\Phi}{dh}}=}
+\frac{1}{2}\,\theta^{ij}\,\partial_j\Phi\star_{h} B_i- \frac{i}{4}\,\theta^{ij}\,\Phi\star_{h} B_j\star_{h} B_i-\frac{i}{2}\,\theta^{ij}\,A_i\star_{h}\Phi\star_{h} B_j}\\[8pt]
{\Phi[a_\rho,b_\rho,\phi;h\theta]\Big{|}_{h=0}=\phi,}
\end{array}
\label{hybridevol}
\end{equation}
where $A_i$ and $B_i$ solve the following equations
\begin{equation}
\begin{array}{c}
{{\displaystyle \frac{d A_{\mu}}{dh}}=\frac{1}{4}\,\theta^{ij}\{A_{i},\partial_{j}A_{\mu}+A_{j\mu}\}_{\star_{h}},\quad A_{\mu}[a_\rho;h\theta]\Big{|}_{h=0}=\,a_{\mu},}\\[8pt]
{{\displaystyle \frac{d B_{\mu}}{dh}}=\frac{1}{4}\,\theta^{ij}\{B_{i},\partial_{j}B_{\mu}+B_{j\mu}\}_{\star_{h}},\quad B_{\mu}[a_\rho;h\theta]\Big{|}_{h=0}=\,b_{\mu},}
\end{array}
\label{swproblems}
\end{equation}
respectively. We use the following notation: $A_{\mu\nu}=\partial_\mu A_\nu-\partial_\nu A_\mu +i[A_\mu,A_\nu]_{\star_{h}}$ and  $B_{\mu\nu}=\partial_\mu B_\nu-\partial_\nu B_\mu +i[B_\mu,B_\nu]_{\star_{h}}$. It has already been shown --see ~\cite{Barnich:2001mc, Barnich:2002pb}-- that (\ref{swproblems}) solve the Seiberg-Witten equations in (\ref{Asweq}) and (\ref{Bsweq}).

To show that by solving  (\ref{hybridevol}) one obtains a hybrid Seiberg-Witten map, we shall take advantage of the cohomological techniques that were developed in Refs.~\cite{Barnich:2001mc, Barnich:2002pb,Barnich:2002tz, Barnich:2003wq} in the context of the antifield formalism. Following ref.~\cite{Barnich:2002pb} we shall prove first that the previous statement is correct for the case of ordinary fields $a_\mu$ and $\lambda$ that take values in the fundamental representation of the Lie algebra of $U(n_L)$, along with ordinary fields $b_\mu$ and $\omega$ which take values in the
fundamental representation of $U(n_R)$. Once the proof for this $(U(n_L), U(n_R))$ case is completed, one finishes the proof for the $(G_L, G_R)$ case by constraining $a_\mu$ and $\lambda$ to take values in the initial $n_L$-dimensional matrix representation of the Lie algebra of $G_L$, and $b_\mu$ and $\omega$ to be valued on the $n_R$-matrix representation of the $R_a$ we started with. Notice that this procedure works --see ref.~\cite{Barnich:2002pb}-- since we are considering faithful representations of the compact Lie algebras of $G_L$ and $G_R$  by hermitian matrices of finite dimension. Hence, until otherwise stated $L_a$ and $R_a$ will be in the fundamental representation of $U(n_L)$ and $U(n_R)$, respectively. This implies that until we say otherwise $a_{\mu}$, $\lambda$, $A_{\mu}$ and $\Lambda$ will be elements of the Lie algebra of $U(n_L)$, with coordinates $a^{a}_{\mu}$, $\lambda^a$, $A^a_{\mu}$ and $\Lambda^a$; and $b_{\mu}$, $\omega$, $B_{\mu}$ and $\Omega$ will be elements of the Lie algebra of $U(n_R)$. with coordinates $b^a_{\mu}$, $\omega^a$, $B^a_{\mu}$ and $\Omega^a$. We should like to point out that the requirement of faithfulness of the representation is a technical condition, not a fundamental one, needed for the approach used here to work.

In the antifield formalism --see~\cite{Henneaux:1992ig, Gomis:1994he}, for a reviews-- one starts by associating an antifield to each field and, then, one sets up the antibracket and the master equation. Let $F^{M}=(A^a_\mu,\Lambda^a,B^a_\mu,\Omega^a,\Phi^{i_L}_{i_R}, \bar{\Phi}^{i_R}_{i_L})$ denote the noncommutative fields collectively. Then $F^{*}_M= (A^{*\,\mu}_a,\Lambda^{*}_a,B^{*\,\mu}_a,\Omega^{*}_a,\Phi^{*\,i_R}_{\phantom{*\,}i_L}, \bar{\Phi}^{*\,i_L}_{\phantom{*\,}i_R})$ will stand for the corresponding noncommutative antifields. Analogously, we have
$f^{M}=(a^a_\mu,\lambda^a,b^a_\mu,\omega^a,\phi^{i_L}_{i_R}, \bar{\phi}^{i_R}_{i_L})$, for the ordinary fields, and $f^{*}_M= (a^{*\,\mu}_a,\lambda^{*}_a,b^{*\,\mu}_a,\omega^{*}_a,\phi^{*\,i_L}_{\phantom{*\,}i_R}, \bar{\phi}^{*\,i_R}_{\phantom{*\,}i_L})$, for the ordinary antifields. The antibracket for  the $F^M$ and $F^{*}_M$ pairs, on the one hand, and $f^M$ and $f^{*}_M$ pairs, on the other, are defined as follows
\begin{equation}
(X,Y)=\idx\,\frac{\partial_r \hat{X}}{\partial F^M}\frac{\partial_l \hat{Y}}{\partial F^{*}_M}-\frac{\partial_r \hat{X}}{\partial F^{*}_M}\frac{\partial_l \hat{Y}}{\partial F^M},\,\quad
(X,Y)=\idx\,\frac{\partial_r X}{\partial f^M}\frac{\partial_l Y}{\partial f^{*}_M}-\frac{\partial_r X}{\partial f^{*}_M}\frac{\partial_l Y}{\partial f^M}.
\label{antibracket}
\end{equation}

The outcome of the analysis carried out in refs~\cite{Barnich:2001mc, Barnich:2002pb, Barnich:2002tz, Barnich:2003wq} is that there are at least three equivalent ways to characterize a Seiberg-Witten map. The way to characterize a Seiberg-Witten map that suits our purposes goes as follows:

A map $F^M[f^{M^{'}},f^{*}_{M^{'}}; h\theta]$,  $F^{*}_M[f^{M^{'}}, f^{*}_{M^{'}};h\theta]$ is a Seiberg-Witten map if, only if, it solves the following problem
\begin{equation}
\begin{array}{l}
{{\displaystyle \frac{d F^M}{dh}}=(\hat{\cal J},F^M),\quad F^M[f^{M^{'}},f^{*}_{M^{'}};h\theta]\Big{|}_{h=0}=f^M,}\\[8pt]
{{\displaystyle \frac{d F^{*}_M}{dh}}=(\hat{\cal J},F^{*}_M),\quad F^{*}_M[f^{M^{'}},f^{*}_{M^{'}};h\theta]\Big{|}_{h=0}=f^{*}_M},
\end{array}
\label{FSWeq}
\end{equation}
where the functional $\hat{\cal J}[F^M,F^{*}_M;h\theta]$  is such that the following equation holds
\begin{equation}
\frac{\partial \hat{S}}{\partial h}= \hat{\cal B}_0\,+\,(\hat{\cal J},\hat{S}),
\label{Jequation}
\end{equation}
for some functional $\hat{\cal B}_0[f^M;h\theta]$, which does not depend on the ordinary antifields $f^{*}_M$. In the previous equation the functional $\hat{S}[F^M,F^{*}_M;h\theta]$ is the minimal proper solution --see Refs.~\cite{Henneaux:1992ig, Gomis:1994he}, for terminology-- of the classical master equation,
\begin{equation}
(\hat{S},\hat{S})=0,
\label{mastereq}
\end{equation}
of the noncommutative gauge theory. In the previous equation the antibracket is defined with regard to the noncommutative fields and antifields --see (\ref{antibracket}).

It is assumed that the functionals $\hat{S}$, $\hat{\cal B}_{0}$ and $\hat{\cal J}$ are polynomials with regard to the star product of the noncommutative fields, noncommutative antifields and their derivatives. This will not be so if we expressed them in terms of the ordinary fields and ordinary antifields.

Let $\hat{S}_{0}[F^M;h\theta]$ denote a real functional which is invariant under the BRST transformations in (\ref{ABRST}), (\ref{BBRST}) and (\ref{HBRST}). $\hat{S}_{0}[F^M;h\theta]$ is the classical noncommutative action of the theory and it is constructed by using the noncommutative field strengths and noncommutative covariant derivatives. An example of such action which is a sum of integrated monomials  of the noncommutative fields, and their derivatives, with mass dimension less than o equal to 4 are given in Appendix A.

It is not difficult to show that
the minimal proper solution, $\hat{S}[F^M,F^{*}_M;h\theta]$, to the master equation (\ref{mastereq}), which satisfies the boundary conditions
\begin{equation*}
\hat{S}[F^M,F^{*}_M=0;h\theta]=\hat{S}_{0}[F^M;h\theta],\quad
\frac{\partial_l \hat{S}}{\partial F^{*}_M}\Big{|}_{F^{*}_{M}=0}=s_{NC}F^M
\end{equation*}
reads
\begin{equation}
\begin{array}{l}
{\hat{S}[F^M,F^{*}_M;h\theta]=\hat{S}_{0}[F^M;h\theta]+\hat{S}_{Antifields}[F^M,F^{*}_M;h\theta],}\\[8pt]
{\hat{S}_{Antifields}[F^M,F^{*}_M;h\theta]=\idx\,\Big(A^{*\,\mu}_a (D_\mu\Lambda)^a+B^{*\,\mu}_a (D_\mu \Omega)^a-i\Lambda^{*}_a(\Lambda\star_h\Lambda)^a-i\Omega^{*}_a(\Omega\star_h\Omega)^a}\\[8pt]
{\phantom{\hat{S}_{Antifields}=\idx}+\Phi^{*\,i_R}_{\phantom{*\,}i_L}(-i\Lambda\star_h\Phi+i\Phi\star_h\Omega)^{i_L}_{i_R}+
\bar{\Phi}^{*\,i_L}_{\phantom{*\,}i_R}(i\bar{\Phi}\star_h\Lambda-i\Omega\star_h\bar{\Phi})^{i_R}_{i_L}\Big).}
\end{array}
\label{masterS}
\end{equation}
Let us recall that, for the time being, the noncommutative fields $A_\mu$ and $\Lambda$ --and their antifields-- take values in the Lie algebra of $U(n_L)$ in the fundamental representation; whereas the noncommutative fields $B_\mu$ and $\Omega$ --and their antifields-
take values in the Lie algebra of $U(n_R)$ in the fundamental representation.

Furnished with  $\hat{S}[F^M,F^{*}_M;h\theta]$ in (\ref{masterS}), we shall look for  a functional $\hat{\cal J}[F^M,F^{*}_M;h\theta]$ such that (\ref{Jequation}) holds. We claim that the  $\hat{\cal J}[F^M,F^{*}_M;h\theta]$ in question reads thus
\begin{equation}
\begin{array}{l}
{\hat{\cal J}[F^M,F^{*}_M;h\theta]=-\idx\;\Big[ A^{*\,\mu}_a\frac{\theta^{ij}}{4}\big(\{A_i,\partial_j A_\mu+A_{j\mu}\}_{\star_h}\big)^a+
B^{*\,\mu}_a\frac{\theta^{ij}}{4}\big(\{B_i,\partial_j B_\mu+B_{j\mu}\}_{\star_h}\big)^a}\\[8pt]
{\phantom{\hat{\cal J}[F^M,F^{*}_M;h\theta]=-\idx\;\Big[}
+\Lambda^{*}_a\frac{\theta^{ij}}{4}\big(\{\partial_i\Lambda,A_j\}_{\star_h}\big)^a+
\Omega^{*}_a\frac{\theta^{ij}}{4}\big(\{\partial_i\Omega,B_j\}_{\star_h}\big)^a}\\[8pt]
{-\Phi^{*\,i_R}_{\phantom{*\,}i_L}\big(\frac{\theta^{ij}}{2}A_i\star_h\partial_j\Phi\!+\!i\frac{\theta^{ij}}{4}
A_i\star_h A_j\star_h\Phi\!+\!\frac{\theta^{ij}}{2}\partial_j\Phi\star_{h}B_i
\!-\!i\frac{\theta^{ij}}{4}\Phi\star_h B_j\star_{h}B_i\!-\!i\frac{\theta^{ij}}{2}A_i\star_h\Phi\star_{h}B_j\big)^{i_L}_{i_R}}\\[8pt]
{-\bar{\Phi}^{*\,i_L}_{\phantom{*\,}i_R}\big(\frac{\theta^{ij}}{2}\partial_j\bar{\Phi}\star_h A_i\!-\!i\frac{\theta^{ij}}{4}
\bar{\Phi}\star_{h}A_j\star_h A_i\!+\!\frac{\theta^{ij}}{2}B_i\star_{h}\partial_j\bar{\Phi}
\!+\!i\frac{\theta^{ij}}{4}B_i\star_{h}B_j\star_{h}\bar{\Phi}\!+\!i\frac{\theta^{ij}}{2}B_j\star_h\bar{\Phi}\star_{h}A_i\big)^{i_R}_{i_L}\Big].}
\end{array}
\label{Jguess}
\end{equation}

Since $\hat{\cal J}$ is linear in the noncommutative antifields $F^{*}_M$, to show that our claim is correct it is enough to prove that the $F^{*}_M$-dependent bit of
\begin{equation*}
\frac{\partial \hat{S}[F^M,F^{*}_{M};h\theta]}{\partial h}
\end{equation*}
is equal to the $F^{*}_M$-dependent part of
\begin{equation*}
(\hat{\cal J },\hat{S}).
\end{equation*}

Let $\hat{\cal A}[F^M,F^{*}_M;h\theta]$ denote the contribution to $(\hat{\cal J },\hat{S})$ which does depend on the noncommutative antifields, $F^{*}_M$, ie, the contribution that vanishes when the noncommutative antifields are set to zero. Now, the fact that $\hat{\cal J}$ is linear in the noncommutative antifields $F^{*}_M$ leads to the conclusion that the classical noncommutative action, $\hat{S}_0[F^M;h\theta]$ --which in turn does not depend on the noncommutative antifields, does not contribute to $\hat{\cal A}[F^M,F^{*}_M;h\theta]$. Indeed,
\begin{equation}
\hat{\cal A}[F^M,F^{*}_M;h\theta]= (\hat{\cal J },\hat{S}_{Antifields}),
\label{defantibit}
\end{equation}
where $\hat{S}_{Antifields}$ is given in (\ref{masterS}). A very long, but straightforward, computation --see Appendix B, for details-- yields the following result:
\begin{equation}
\begin{array}{l}
{\hat{\cal A}[F^M,F^{*}_M;h\theta]=-\frac{\theta^{ij}}{2}\,\idx\,\Big[A^{*\,\mu}_a (\{\partial_i A_\mu,\partial_j\Lambda\}_{\star_h})^a+B^{*\,\mu}_a (\{\partial_i B_\mu,\partial_j\Omega\}_{\star_h})^a}\\[8pt]
{\phantom{\hat{\cal A}[F^M,F^{*}_M;h\theta]=i\frac{\theta^{ij}}{2}\,\idx\,\Big[}
-\Lambda^{*}_a(\partial_i\Lambda\star_h\partial_j\Lambda)^a+\Omega^{*}_a(\partial_i\Omega\star_h\partial_j\Omega)^a}\\[8pt]
{\phantom{\hat{\cal A}[F^M,F^{*}_M;h}
+\Phi^{*\,i_R}_{\phantom{*\,}i_L}(-\partial_i\Lambda\star_h\partial_j\Phi+\partial_i\Phi\star_h\partial_j\Omega)^{i_L}_{i_R}+
\bar{\Phi}^{*\,i_L}_{\phantom{*\,}i_R}(\partial_i\bar{\Phi}\star_h\partial_j\Lambda-\partial_i\Omega\star_h\partial_j\bar{\Phi})^{i_R}_{i_L}\Big].}
\end{array}
\label{antibit}
\end{equation}

By computing the partial derivative of $\hat{S}_{Antifields}[F^M,F^{*}_M;h\theta]$ in (\ref{masterS}) with respect to $h$ --recall that no derivatives of $F^M$ and $F^{*}_M$ with respect to $h$ are taken, one also obtains
the R.H.S of (\ref{antibit}). Thus we come to be conclusion that
\begin{equation*}
\begin{array}{l}
{\frac{\partial \hat{S}[F^M,F^{*}_{M};h\theta]}{\partial h}-(\hat{\cal J },\hat{S})=
\frac{\partial \hat{S}_0[F^M;h\theta]}{\partial h}-(\hat{\cal J },\hat{S}_{0})+\frac{\partial \hat{S}_{Antifields}[F^M,F^{*}_{M};h\theta]}{\partial h}-\hat{\cal A}[F^M,F^{*}_M;h\theta]}\\[8pt]
{\phantom{\frac{\partial \hat{S}[F^M,F^{*}_{M};h\theta]}{\partial h}-(\hat{\cal J },\hat{S})}=\frac{\partial \hat{S}_0[F^M;h\theta]}{\partial h}-(\hat{\cal J },\hat{S}_{0})\,=\,
{\cal B}_0[A^a_\mu,B^a_\mu,\Phi^{i_L}_{i_R},\hat{\Phi}^{i_R}_{i_L};h\theta].}
\end{array}
\end{equation*}
It is key to realize that ${\cal B}_0[A^a_\mu,B^a_\mu,\Phi^{i_L}_{i_R},\hat{\Phi}^{i_R}_{i_L};h\theta]$ does not depend on the noncommutative antifields.

Now, taking into account that ${\hat{\cal J}}$ in (\ref{Jguess}) is linear in the noncommutative antifields, one comes to the conclusion that $(\hat{{\cal J}},F^M)$ does not depend on the noncommutative antifields. Hence the solution to the
``evolution'' problem
\begin{equation}
 {{\displaystyle \frac{d F^M}{dh}}=(\hat{\cal J},F^M),\quad F^M[f^{M^{'}},f^{*}_{M^{'}};h\theta]\Big{|}_{h=0}=f^M}
 \label{swcalJ}
\end{equation}
only involves the ordinary fields, $f^M$, and not the ordinary antifields $f^{*}_{M}$: $F^{M}=F^M[f^{M^{'}};h\theta]$. Thus, in our case ${\cal B}_0[A^a_\mu,B^a_\mu,\Phi^{i_L}_{i_R},\hat{\Phi}^{i_R}_{i_L};h\theta]$  does not depend on the
ordinary antifields when we replace $A^a_\mu$, $B^a_\mu$, $\Phi^{i_L}_{i_R}$ and $\hat{\Phi}^{i_R}_{i_L}$ in (\ref{Jguess}) with the corresponding solution to (\ref{swcalJ}). We have thus finished the proof that the equations in (\ref{FSWeq}) define a Seiberg-Witten map for the $\hat{\cal J}$ in (\ref{Jguess}).

Notice that for $\hat{\cal J}$ in (\ref{Jguess}), one has
\begin{equation}
\begin{array}{l}
{(\hat{\cal J},A^a_\mu)L_a= \frac{1}{4}\,\theta^{ij}\{A_{i},\partial_{j}A_{\mu}+A_{j\mu}\}_{\star_{h}},\quad
(\hat{\cal J},\Lambda^a)L_a=\frac{1}{4}\,\theta^{ij}\{\partial_i  \Lambda,A_j\}_{\star_{h}},}\\[8pt]
{(\hat{\cal J},B^a_\mu)R_a= \frac{1}{4}\,\theta^{ij}\{B_{i},\partial_{j}B_{\mu}+B_{j\mu}\}_{\star_{h}},\quad
(\hat{\cal J},\Omega^a)R_a=\frac{1}{4}\,\theta^{ij}\{\partial_i\Omega,B_j\}_{\star_{h}},}\\[8pt]
{(\hat{\cal J},\Phi^{i_L}_{i_R})=\big(\frac{1}{2}\,\theta^{ij}A_i\star_{h}\partial_j\Phi
+\frac{i}{4}\,\theta^{ij}\,A_i\star_{h} A_j\star_{h}\Phi\big)^{i_L}_{i_R}}\\[8pt]
{\phantom{(\hat{\cal J},\Phi^{i_L}_{i_R})=\big(\frac{1}{2}\,\quad\quad\quad\quad\quad\quad}
+\big(\frac{1}{2}\,\theta^{ij}\,\partial_j\Phi\star_{h} B_i- \frac{i}{4}\,\theta^{ij}\,\Phi\star_{h} B_j\star_{h} B_i-\frac{i}{2}\,\theta^{ij}\,A_i\star_{h}\Phi\star_{h} B_j\big)^{i_L}_{i_R},}\\[8pt]
{(\hat{\cal J},\bar{\Phi}^{i_R}_{i_L})=
\big(\frac{1}{2}\,\theta^{ij}\,\partial_j\bar{\Phi}\star_{h} A_i-\frac{i}{4}\,\theta^{ij}\,\bar{\Phi}\star_h A_j\star_{h} A_i\big)^{i_R}_{i_L}}\\[8pt]
{\phantom{(\hat{\cal J},\Phi^{i_L}_{i_R})=\big(\frac{1}{2}\,\quad\quad\quad\quad\quad\quad}
+\big(\frac{1}{2}\,\theta^{ij}\,B_i\star_h\partial_j\bar{\Phi}+ \frac{i}{4}\,\theta^{ij}\, B_i\star_{h} B_j\star_{h}\bar{\Phi}+\frac{i}{2}\,\theta^{ij}\,B_j\star_{h}\bar{\Phi}\star_{h} A_i\big)^{i_R}_{i_L},}
\end{array}
\label{antibracketsw}
\end{equation}
where $L_a$ and $R_a$ are the generators of $U(n_L)$ and $U(n_R)$ in the corresponding fundamental representations. $L_a$ and $R_a$ are normalized so that $Tr(L_a L_b)=\delta_{ab}$ and $Tr(R_a R_b)=\delta_{ab}$. Hence,
taking into account the results in (\ref{antibracketsw}) and the equations in (\ref{swcalJ}), one concludes that the ``evolution" equations in (\ref{hybridevol}) and (\ref{swproblems}) define a Seiberg-Witten map.

So far the ordinary fields $a_\mu$ and $\lambda$ take values in the Lie algebra of $U(n_L)$, in the fundamental representation, and  the ordinary fields $b_\mu$ and $\omega$ take values in Lie algebra of $U(n_R)$, also in the fundamental representation. Let us now move on and consider the case when the ordinary gauge fields and ghosts take values in faithful matrix representations of Lie algebras of compact Lie groups.

Let  ${\mathcal M}_L$ denote the Lie algebra of $n_L\times n_L$ matrices which constitutes the finite faithful representation of the Lie algebra of the compact Lie group $G_L$ we had at the beginning of this section. Analogously, let ${\mathcal M}_R$ denote the Lie algebra of $n_R\times n_R$ matrices which realize a faithful representation of the Lie algebra of the compact Lie group $G_R$ we introduced above. ${\mathcal M}_L$ is a Lie subalgebra of the Lie algebra of $U(n_L)$ in the fundamental representation. Similarly,  ${\mathcal M}_R$ is a Lie subalgebras of the Lie algebra of $U(n_R)$ in the fundamental representation. Then, then by restricting $a_\mu$ and $\lambda$ to take values in  ${\mathcal M}_L$, and $b_\mu$ and $\omega$ to take values in  ${ \mathcal M}_R$, we conclude that the ``evolution" equations in (\ref{hybridevol}) and (\ref{swproblems}) define a hybrid Seiberg-Witten map for arbitrary compact groups in faithful unitary finite dimensional representations.

\section{Solving the hybrid Seiberg-Witten map equation in a $\theta$-exact way}

Let us embrace the notion that in a noncommutative quantum field theory each interaction vertex in momentum space is a monomial in the ordinary fields. Then one finds it  natural  to solve the problem in (\ref{hybridevol}) by expanding $\Phi[a_\mu,b_\mu,\phi; h\theta]$ in the number of ordinary gauge fields. Hence,
$\Phi[a_\mu, b_\mu,\phi;h\theta]$ will be given by
\begin{equation}
\Phi[a_\mu, b_\mu,\phi;h\theta]=\displaystyle\sum_{n\geq 0}\;\Phi^{(n)}[a_\mu,b_\mu,\phi;h\theta],
\label{Phiexpan}
\end{equation}
where the superscript $n$ in  $\Phi^{(n)}[a_\mu,b_\mu,\phi;h\theta]$ signals that its Fourier transform is a monomial of degree $n$ in the ordinary gauge fields.  Obviously,
\begin{equation}
\Phi^{(0)}[a_\mu,b_\mu,\phi;h\theta]\big{|}_{h=0}=\phi,\quad n>0\implies\Phi^{(n)}[a_\mu,b_\mu,\phi;h\theta]\big{|}_{h=0}=0,
\label{initialcon}
\end{equation}
if the ``initial" condition in (\ref{hybridevol}) is to be met.

Substituting the expansion in (\ref{Phiexpan}) in the ``evolution" equation in (\ref{hybridevol}), one finds that the differential equation can be solved recursively. Indeed,  $\Phi^{(n)}[a_\mu,b_\mu,\phi;h\theta]$ is given by
\begin{equation*}
\begin{array}{l}
{{\displaystyle \frac{d\Phi^{(n)}}{dh}}=\frac{1}{2}\,\theta^{ij}\displaystyle\sum_{m_1+m_2=n}\,A^{(m_2)}_i\star_{h}\partial_j\Phi^{(m_1)}+\frac{i}{4}\,\theta^{ij}\displaystyle\sum_{m_1+m_2+m_3=n}\,A_i^{(m_2)}\star_{h} A_j^{(m_3)}\star_{h}\Phi^{(m_1)}}\\[8pt]
{\phantom{{\displaystyle \frac{d\Phi}{dh}}=\quad}
+\frac{1}{2}\,\theta^{ij}\displaystyle\sum_{m_1+m_2=n}\,\partial_j\Phi^{(m_1)}\star_{h} B^{(m_2)}_i- \frac{i}{4}\,\theta^{ij}\,\displaystyle\sum_{m_1+m_2+m_3=n}\Phi^{(m_1)}\star_{h} B^{(m_3)}_j\star_{h} B^{(m_2)}_i}\\[8pt]
{\phantom{{\displaystyle \frac{d\Phi}{dh}}=\quad}
-\frac{i}{2}\,\theta^{ij}\displaystyle\sum_{m_1+m_2+m_3=n}\,A^{(m_2)}_i\star_{h}\Phi^{(m_1)}\star_{h} B^{(m_3)}_j}.
\end{array}
\end{equation*}
It is important to stress that in the previous equation $m_2\geq 1$ and $m_3 \geq 1$, whereas $m_1\geq 0$. $A^{(m)}_\mu[a_\nu;h\theta]$ and $B^{(m)}_\mu[b_\nu;h\theta]$ are such that their Fourier transform are monomials of degree $m$ in $a_\nu$ and
$b_\nu$, respectively,  and they furnish the following solutions to the Seiberg-Witten problems in (\ref{swproblems}):
\begin{equation*}
A_\mu[a_\mu;h\theta]=\displaystyle\sum_{m\geq 1}\;\A^{(m)}_\mu[a_\nu;h\theta],\quad B_\mu[b_\nu;h\theta]=\displaystyle\sum_{m\geq 1}\;\B^{(m)}_\mu[b_\nu;h\theta].
\end{equation*}
$A^{(m)}_\mu[a_\nu;h\theta]$ --and, therefore $B^{(m)}_\mu[b_\nu;h\theta]$-- has been computed in  ~\cite{Martin:2012aw} for $m=1,2,3$.

Let us work out $\Phi^{(n)}[a_\mu,b_\mu,\phi;h\theta]$ for $n=0,1,2,3$. The equations to be solved recursively, for the ``initial" conditions in (\ref{initialcon}), read
\begin{equation}
\begin{array}{l}
{{\displaystyle\frac{ d\Phi^{(0)}}{dh}}=0, }\\[12pt]
{{\displaystyle\frac{d \Phi^{(1)}}{dh}}=\frac{1}{2}\,\theta^{ij}A^{(1)}_{i}\star_{h}\partial_{j}\Phi^{(0)}+\frac{1}{2}\,\theta^{ij}\partial_{j}\Phi^{(0)}\star_{h} B^{(1)}_{i}, }\\[12pt]
{{\displaystyle\frac{d \Phi^{(2)}}{dh}}=\phantom{+}\frac{1}{2}\,\theta^{ij}A^{(1)}_{i}\star_{h}\partial_{j}\Phi^{(1)}+
\frac{1}{2}\,\theta^{ij}A^{(2)}_{i}\star_{h}\partial_{j}\Phi^{(0)}+\frac{i}{4}\,\theta^{ij}\,A^{(1)}_i\star_h A^{(1)}_j\star_h \Phi^{(0)}}\\[12pt]
{\phantom{{\displaystyle\frac{d \Phi^{(2)}}{dh}}=}
{+\frac{1}{2}\,\theta^{ij}\partial_{j}\Phi^{(1)}\star_{h}}B^{(1)}_{i}+
\frac{1}{2}\,\theta^{ij}\partial_{j}\Phi^{(0)}\star_{h}B^{(2)}_{i}
-\frac{i}{4}\,\theta^{ij}\,\Phi^{(0)}\star_h B^{(1)}_j\star_h B^{(1)}_i }\\[12pt]
{\phantom{{\displaystyle\frac{d \Phi^{(2)}}{dh}}=}
-\frac{i}{2}\,\theta^{ij}\,A^{(1)}_{i}\star_{h}\Phi^{(0)}\star_h B^{(1)}_j ,}\\[12pt]
{{\displaystyle\frac{d \Phi^{(3)}}{dh}}=\phantom{+}\frac{1}{2}\,\theta^{ij}A^{(3)}_{i}\star_{h}\partial_{j}\Phi^{(0)}+
\frac{1}{2}\,\theta^{ij}A^{(2)}_{i}\star_{h}\partial_{j}\Phi^{(1)}+ \frac{1}{2}\,\theta^{ij}A^{(1)}_{i}\star_{h}\partial_{j}\Phi^{(2)} }\\[10pt]
{\phantom{{\displaystyle\frac{d \Phi^{(3)}}{dh}}=}+\frac{i}{4}\,\theta^{ij}\,A^{(2)}_i\star_h A^{(1)}_j\star_h \Phi^{(0)}+\frac{i}{4}\,\theta^{ij}\,A^{(1)}_i\star_h A^{(2)}_j\star_h \Phi^{(0)}+\frac{i}{4}\,\theta^{ij}\,A^{(1)}_i\star_h A^{(1)}_j\star_h \Phi^{(1)}}\\[12pt]
{\phantom{{\displaystyle\frac{d \Phi^{(3)}}{dh}}=}
+\frac{1}{2}\,\theta^{ij}\partial_{j}\Phi^{(0)}\star_{h}B^{(3)}_{i}+\frac{1}{2}\,\theta^{ij}\partial_{j}\Phi^{(1)}\star_{h} B^{(2)}_{i}+ \frac{1}{2}\,\theta^{ij}\partial_{j}\Phi^{(2)}\star_h B^{(1)}_{i} }\\[10pt]
{\phantom{{\displaystyle\frac{d \Phi^{(3)}}{dh}}=}
-\frac{i}{4}\,\theta^{ij}\,\Phi^{(0)}\star_h B^{(1)}_j\star_h B^{(2)}_i-\frac{i}{4}\,\theta^{ij}\,\Phi^{(0)}\star_h B^{(2)}_j\star_h B^{(1)}_i-\frac{i}{4}\,\theta^{ij}\,\Phi^{(1)}\star_h B^{(1)}_j \star_h  B^{(1)}_i }\\[12pt]
{\phantom{{\displaystyle\frac{d \Phi^{(3)}}{dh}}=}
-\frac{i}{2}\,\theta^{ij}\,A^{(2)}_i\star_h\Phi^{(0)}\star_h B^{(1)}_j-\frac{i}{2}\,\theta^{ij}\,A^{(1)}_{i}\star_{h}\Phi^{(1)}\star_h B^{(1)}_j-\frac{i}{2}\theta^{ij}\,A^{(1)}_i\star_h\Phi^{(0)}\star_h B^{(2)}_j . }
\end{array}
\label{recursivephieqs}
\end{equation}
Hence, by integrating with regard to $h$ both sides of each differential equation in (\ref{recursivephieqs}), one obtains
\begin{equation}
\begin{array}{l}
{ \Phi^{(0)}[a_\mu,b_\mu,\phi;h\theta]=\phi, }\\[12pt]
{\Phi^{(1)}[a_\mu,b_\mu,\phi;h\theta]=\int_{0}^{h}\,dt\,\Big(\frac{1}{2}\theta^{ij}a_{i}\star_{t}\partial_{j}\phi+\frac{1}{2}\theta^{ij}\partial_{j}\phi\star_t b_i\Big), }\\[12pt]
{\Phi^{(2)}[a_\rho,b_\rho,\Phi;h\theta]=\int_{0}^{h}\,dt\,\Big(\phantom{+}\frac{1}{2}\theta^{ij}a_{i}\star_{t}\partial_{j}\Phi^{(1)}[t\theta]+
\frac{1}{2}\theta^{ij}A^{(2)}_i[t\theta]\star_t\partial_{j}\phi+\frac{i}{4}\theta^{ij}\,a_i\star_t a_j\star_t \phi}\\[12pt]
{\phantom{\Phi^{(2)}[a_\rho,\Phi;h\theta]=\int_{0}^{h}\,dt\,\Big(}
+\frac{1}{2}\theta^{ij}\partial_{j}\Phi^{(1)}[t\theta]\star_{t}b_{i}+
\frac{1}{2}\theta^{ij}\partial_{j}\phi\star_t B^{(2)}_i[t\theta]-\frac{i}{4}\theta^{ij}\phi\star_t b_j\star_t b_i}\\[12pt]
{\phantom{\Phi^{(2)}[a_\rho,\Phi;h\theta]=\int_{0}^{h}\,dt\,\Big(}
-\frac{i}{2}\theta^{ij}\,a_i\star_t\phi\star_t b_j\Big),}\\[12pt]
{\Phi^{(3)}[a_\mu,b_\mu,\phi;h\theta]=\int_{0}^{h}\,dt\,}\\[12pt]
{\quad\quad\Big(\phantom{+}\frac{1}{2}\theta^{ij}A^{(3)}_{i}[t\theta]\star_{t}\partial_{j}\phi+\frac{1}{2}\theta^{ij}A^{(2)}_{i}[t\theta]\star_{t}\partial_{j}\Phi^{(1)}[t\theta]
+\frac{1}{2}\,\theta^{ij}a_{i}\star_{t}\partial_{j}\Phi^{(2)}[t\theta] }\\[10pt]
{\quad\quad+\frac{i}{4}\,\theta^{ij}\,A^{(2)}_i[t\theta]\star_t a_j\star_t \phi+\frac{i}{4}\,\theta^{ij}\,a_i\star_t A^{(2)}_j[t\theta]\star_t \phi+\frac{i}{4}\,\theta^{ij}\,a_i\star_t a_j\star_t \Phi^{(1)}[t\theta]}\\[12pt]
{\quad\quad+\frac{1}{2}\,\theta^{ij}\partial_{j}\phi\star_{t}B^{(3)}_{i}[t\theta]+\frac{1}{2}\theta^{ij}\partial_{j}\Phi^{(1)}[t\theta]\star_{t} B^{(2)}_{i}[t\theta]+ \frac{1}{2}\theta^{ij}\partial_{j}\Phi^{(2)}[t\theta]\star_t b_{i}[t\theta] }\\[10pt]
{\quad\quad
-\frac{i}{4}\,\theta^{ij}\phi\star_t b_j[t\theta]\star_t B^{(2)}_i[t\theta]-\frac{i}{4}\theta^{ij}\phi\star_t B^{(2)}_j[t\theta]\star_t b_i[t\theta]-\frac{i}{4}\theta^{ij}\Phi^{(1)}[t\theta]\star_t b_j \star_t  b_i }\\[12pt]
{\quad\quad
-\frac{i}{2}\,\theta^{ij}A^{(2)}_i[t\theta]\star_t\phi\star_t b_j[t\theta]-\frac{i}{2}\theta^{ij} a_{i}[t\theta]\star_{t}\Phi^{(1)}[t\theta]\star_t b_j[t\theta]-\frac{i}{2}\theta^{ij}\,a_i\star_t\phi\star_t B^{(2)}_j[t\theta]\Big), }
\end{array}
\label{iterativesolpsi}
\end{equation}
where we have taken into account that $A^{(1)}[a_\mu;h\theta]=a_\mu$ and $B^{(1)}[b_\mu;h\theta]=b_\mu$ --see~\cite{Martin:2012aw}.

Next, let us carry out the integrations over $t$ in the  integrals in (\ref{iterativesolpsi}). Then, the following expressions for  $\Phi^{(1)}$ and $\Phi^{(2)}$ are obtained in momentum space:
\begin{equation*}
\begin{array}{l}
{\Phi^{(1)\,i_{L}}_{\phantom{(1)}\,i_{R}}(x)=\phantom{+}\int\frac{{\rm d}^4 p_{1}}{(2\pi)^4}\frac{{\rm d}^4 p_{2}}{(2\pi)^4}e^{-i(p_{1}+p_{2})x}\;\theta^{ij}p_{2j}\frac{e^{-i\frac{h}{2}(p_{1}\wedge p_{2})}-1}{p_{1}\wedge p_{2}}\;(L_a)^{i_L}_{j_{L}}\;a_{i}^{a}(p_{1}){\phi}(p_{2})^{j_{L}}_{i_{R}} }\\[12pt]
{\phantom{\Phi^{(1)\,i_{L}}_{\phantom{(1)}\,i_{R}}(x)=}
+\int\frac{{\rm d}^4 p_{1}}{(2\pi)^4}\frac{{\rm d}^4 p_{2}}{(2\pi)^4}e^{-i(p_{1}+p_{2})x}\;\theta^{ij}p_{2j}\frac{e^{-i\frac{h}{2}(p_{2}\wedge p_{1})}-1}{p_{2}\wedge p_{1}}\;
 (R_b)^{j_R}_{i_{R}}\;b_{i}^{b}(p_{1})}{\phi}(p_{2})^{i_{L}}_{j_{R}},
\end{array}
\end{equation*}
\begin{equation*}
\begin{array}{l}
{\Phi^{(2)\,i_{L}}_{\phantom{(2)}\,i_{R}}(x) =\int\prod_{i=1}^{3}\frac{{\rm d}^4 p_{i}}{(2\pi)^{4}}\;e^{-i(p_{1}+p_{2}+p_{3})x}\;}\\[12pt]
{\phantom{\Phi^{(2)\,i_{L}}_{\phantom{(2)}\,i_{R}}(x) =\int\prod_{i=1}^{3}\frac{{\rm d}^4 p_{i}}{(2\pi)^{4}}}
\{\mathbb{M}^{(2,0)}[(\mu_{1},p_{1});(\mu_{2},p_{2});p_{3};h\theta](L_{a_{1}}L_{a_{2}})^{i_{L}}_{j_{L}}\; a_{\mu_{1}}^{a_{1}}(p_{1})a_{\mu_{2}}^{a_{2}}(p_{2})\phi(p_{3})^{j_{L}}_{i_{R}} }\\[12pt]
{\phantom{\Phi^{(2)\,i_{L}}_{\phantom{(2)}\,i_{R}}(x) =\int\prod_{i=1}^{3}\frac{{\rm d}^4 p_{i}}{(2\pi)^{4}}}
+\mathbb{M}^{(1,1)}[(\mu_1,p_1);(\mu_2,p_2);p_3;h\theta](L_{a_1})^{i_L}_{j_L} (R_{a_2})^{j_R}_{i_R}\; a_{\mu_1}^{a_1}(p_1)b_{\mu_2}^{a_2}(p_2)\phi(p_3)^{j_L}_{j_R} }\\[12pt]
{\phantom{\Phi^{(2)\,i_{L}}_{\phantom{(2)}\,i_{R}}(x) =\int\prod_{i=1}^{3}\frac{{\rm d}^4 p_{i}}{(2\pi)^{4}}}
+\mathbb{M}^{(0,2)}[(\mu_{1},p_{1});(\mu_{2},p_{2});p_{3};h\theta](R_{a_{1}}R_{a_{2}})^{j_{R}}_{i_{R}}\;b_{\mu_{1}}^{a{1}}(p_{1})b_{\mu_{2}}^{a_2}(p_{2})\phi(p_{3})^{i_{L}}_{j_{R}}\}; }
\end{array}
\end{equation*}
where
\begin{equation*}
\begin{array}{l}
{\mathbb{M}^{(2,0)}[(\mu_{1},p_{1});(\mu_{2},p_{2});p_{3};h\theta]= -\frac{1}{2}\,\theta^{ij}\;\delta_i^{\mu_1}\delta_j^{\mu_2}\,
\displaystyle\Big[\frac{e^{-i\frac{h}{2}(p_1\wedge p_2+ p_1\wedge p_3+p_2\wedge p_3)}-1}{ p_1\wedge p_2+ p_1\wedge p_3+p_2\wedge p_3}\Big]}\\[10pt]
{\quad\quad\quad\quad\quad+\theta^{ij}\theta^{kl}\,\delta_i^{\mu_1}\delta_k^{\mu_2}(p_2\!+\!p_3)_j\, p_{3 l}\,
\displaystyle\frac{1}{p_2\wedge p_3}\Big[\frac{e^{-i\frac{h}{2}(p_1\wedge p_2+ p_1\wedge p_3+p_2\wedge p_3)}-1}{ p_1\wedge p_2+ p_1\wedge p_3+p_2\wedge p_3}-
\frac{e^{-i\frac{h}{2} p_1\wedge( p_2+p_3)}-1}{p_1\wedge( p_2+p_3)}\Big]}\\[10pt]
{\quad\quad\quad\quad\quad+\frac{1}{2}\,\theta^{ij}\theta^{kl}\;
[2(p_{2 l}\,\delta_k^{\mu_1}\delta_i^{\mu_2}+p_{1 l}\,\delta_k^{\mu_2}\delta_i^{\mu_1})-(p_2\!-\!p_1)_{i}\,\delta_k^{\mu_1} \delta_l^{\mu_2}]\,p_{3 j}}\\[10pt]
{\phantom{\quad\quad\quad\quad\quad+\theta^{ij}\theta^{kl}\,\delta_i^{\mu_1}\delta_k^{\mu_2}(p_2\!+\!p_3)_j\, p_{3 l}\,}
\displaystyle\frac{1}{p_1\wedge p_2}\Big[\frac{e^{-i\frac{h}{2}(p_1\wedge p_2+ p_1\wedge p_3+p_2\wedge p_3)}-1}{ p_1\wedge p_2+ p_1\wedge p_3+p_2\wedge p_3}-
\frac{e^{-i\frac{h}{2}( p_1+p_2)\wedge p_3}-1}{( p_1+p_2)\wedge p_3}\Big],}\\[12pt]
{\mathbb{M}^{(1,1)}[(\mu_{1},p_{1});(\mu_{2},p_{2});p_{3};h\theta]=\theta^{ij}\theta^{kl}\delta^{\mu_{1}}_{k}\delta^{\mu_{2}}_{i}(p_{1}+p_{3})_{j}p_{3l}}\\[10pt]
{\phantom{\mathbb{B}[(\mu_{1},p_{1});(\mu_{2},p_{2});p_{3};\theta]=\theta^{ij}\theta^{kl}\delta^{\mu_{1}}_{k}}
\displaystyle \frac{1}{p_{1}\wedge p_{3}}\Big[\frac{e^{-i\frac{h}{2}(p_{1}\wedge p_{2}+p_{1}\wedge p_{3}+p_{3}\wedge p_{2})}-1}{p_{1}\wedge p_{2}+p_{1}\wedge p_{3}+p_{3}\wedge p_{2}}-
\frac{e^{-i\frac{h}{2}(p_{1}\wedge p_{2}+p_{3}\wedge p_{2})}-1}{p_{1}\wedge p_{2}+p_{3}\wedge p_{2}}\Big]}\\[10pt]
{\quad\quad-\theta^{ij}\theta^{kl}(p_{2}+p_{3})_{j}p_{3l}\delta_{i}^{\mu_{1}}\delta_{k}^{\mu_{2}}
\displaystyle\frac{1}{p_{2}\wedge p_{3}}\Big[\frac{e^{-i\frac{h}{2}(p_{1}\wedge p_{2}+p_{1}\wedge p_{3}+p_{3}\wedge p_{2})}-1}{p_{1}\wedge p_{2}+p_{1}\wedge p_{3}+p_{3}\wedge p_{2}}-\frac{e^{-i\frac{h}{2}(p_{1}\wedge p_{2}+p_{1}\wedge p_{3})}-1}{p_{1}\wedge p_{2}+p_{1}\wedge p_{3}}\big]}\\[10pt]
{\phantom{\mathbb{B}[(\mu_{1},p_{1});(\mu_{2},p_{2});p_{3};h\theta]=\theta^{ij}\theta^{kl}\delta^{\mu_{1}}_{k}}
+ \theta^{ij}\;\delta_i^{\mu_1}\delta_j^{\mu_2}\displaystyle\frac{e^{-i\frac{h}{2}(p_1\wedge p_2+ p_1\wedge p_3+p_3\wedge p_2)}-1}{ p_1\wedge p_2+ p_1\wedge p_3+p_3\wedge p_2},}\\[14pt]
{\mathbb{M}^{(0,2)}[(\mu_{1},p_{1});(\mu_{2},p_{2});p_{3};h\theta]=\overline{\mathbb{M}^{(2,0)}}[(\mu_{2},-p_{2});(\mu_{1},-p_{1});-p_{3};h\theta].}
\end{array}
\end{equation*}
The bar above $\mathbb{M}^{(2,0)}$ stands for complex conjugate.

 To carry out the integration over $t$ in the expression in (\ref{iterativesolpsi}) giving $\Phi^{(3)}[a_\mu,b_\mu,\phi;h\theta]$, one needs $A^{(3)}_{i}[t\theta]$,
 $A^{(2)}_{i}[t\theta]$, $B^{(3)}_{i}[t\theta]$ and $B^{(2)}_{i}[t\theta]$: these are given in ref.~\cite{Martin:2012aw}. A lengthy computation yields
 \begin{equation}
\begin{array}{l}
{\Phi^{(3)\,i_{L}}_{\phantom{(3)}\,i_{R}}(x) =\int\prod_{i=1}^{4}\frac{{\rm d}^4 p_{i}}{(2\pi)^{4}}\;e^{-i(p_{1}+p_{2}+p_{3}+p_{4})x}\;}\\[12pt]
{\phantom{\Phi^{(3)\,i_{L}}_{\phantom{(3)}\,i_{R}}(x)}
\{\mathbb{M}^{(3,0)}[(\mu_{1},p_{1});(\mu_{2},p_{2});(\mu_{3},p_{3});p_{4};h\theta](L_{a_{1}}L_{a_{2}}L_{a_{3}})^{i_{L}}_{j_{L}}\; a_{\mu_{1}}^{a_{1}}(p_{1})a_{\mu_{2}}^{a_{2}}(p_{2})a_{\mu_{3}}^{a_3}(p_{3})\phi(p_{4})^{j_{L}}_{i_{R}} }\\[12pt]
{\phantom{\Phi^{(3)\,i_{L}}_{\phantom{(3)}\,i_{R}}}
+\mathbb{M}^{(2,1)}[(\mu_1,p_1);(\mu_2,p_2);(\mu_3,p_3);p_4;h\theta](L_{a_1}L_{a_2})^{i_L}_{j_L} (R_{a_3})^{j_R}_{i_R}\; a_{\mu_1}^{a_1}(p_1)a_{\mu_2}^{a_2}(p_2)b_{\mu_3}^{a_3}(p_3)\phi(p_4)^{j_L}_{j_R} }\\[12pt]
{\phantom{\Phi^{(3)\,i_{L}}_{\phantom{(3)}\,i_{R}}}
+\mathbb{M}^{(1,2)}[(\mu_1,p_1);(\mu_2,p_2);(\mu_3,p_3);p_4;h\theta](L_{a_1})^{i_L}_{j_L} (R_{a_2}R_{a_3})^{j_R}_{i_R}\; a_{\mu_1}^{a_1}(p_1)b_{\mu_2}^{a_2}(p_2)b_{\mu_3}^{a_3}(p_3)\phi(p_4)^{j_L}_{j_R} }\\[12pt]
{\phantom{\Phi^{(3)\,i_{L}}_{\phantom{(3)}\,i_{R}}}
+\mathbb{M}^{(0,3)}[(\mu_{1},p_{1});(\mu_{2},p_{2});(\mu_{3},p_{3});p_{4};h\theta](R_{a_{1}}R_{a_{2}}R_{a_{3}})^{i_{L}}_{j_{L}}\; b_{\mu_{1}}^{a_{1}}(p_{1})b_{\mu_{2}}^{a_{2}}(p_{2})b_{\mu_{3}}^{a_3}(p_{3})\phi(p_{4})^{j_{L}}_{i_{R}} \; \},}
\end{array}
\label{3order}
\end{equation}
where $\mathbb{M}^{(3,0)}[\,\cdot\,;\theta]$, $\mathbb{M}^{(2,1)}[\,\cdot\,;\theta]$,  $\mathbb{M}^{(1,2)}[\,\cdot\,;\theta]$ and $\mathbb{M}^{(0,3)}[\,\cdot\,;\theta]$ are given in Appendix C.

\newpage

\section{Hybrid Seiberg-Witten maps of the Higgs field in the noncommutative Standard Model}

In this section, $a_\mu(x)$, $b_\mu(x)$  and $G_\mu(x)$ will denote the $U(1)$, $SU(2)$ and $SU(3)$ gauge fields of the ordinary Standard Model; $\phi(x)$ will stand for the ordinary Higgs doublet and $\unit_2$ will stand for the unit on $\mathbb{C}^2$. Let us recall that $a_\mu(x)$ is a real vector field, that $b_\mu(x)$ is a hermitian complex matrix and that $\phi(x)$ takes values in $\mathbb{C}^2$. Below, we shall use the entries, $G^{\phantom{\mu}\,s_1}_{\mu\,s_2}(x)$, $s_1,s_2=1,2,3$, of the matrix $G_\mu(x)$, rather than the matrix itself, and, thus,  make apparent the doublet structure of the expressions displayed therein.

The reader should look up, in the previous section, the definitions of the functions
$\mathbb{M}^{(2,0)}[(\mu_{1},p_{1});(\mu_{2},p_{2});p_{3};h\theta]$, $\mathbb{M}^{(1,1)}[(\mu_{1},p_{1});(\mu_{2},p_{2});p_{3};h\theta]$, $\mathbb{M}^{(0,2)}[(\mu_{1},p_{1});(\mu_{2},p_{2});p_{3};h\theta]$, $\mathbb{M}^{(2,1)}[(\mu_{1},p_{1});(\mu_{2},p_{2});(\mu_{3},p_{3});p_{4};h\theta]$,
$\mathbb{M}^{(3,0)}[(\mu_{1},p_{1});(\mu_{2},p_{2});(\mu_{3},p_{3});p_{4};h\theta]$,\\
$\mathbb{M}^{(3,0)}[(\mu_{1},p_{1});(\mu_{2},p_{2});(\mu_{3},p_{3});p_{4};h\theta]$ and $\mathbb{M}^{(1,2)}[(\mu_{1},p_{1});(\mu_{2},p_{2});(\mu_{3},p_{3});p_{4};h\theta]$, which
shall occur below.

The construction of the noncommutative Yukawa terms of the noncommutative Standard Model of Ref.~\cite{Calmet:2001na} requires three types of hybrid Seiberg-Witten map of the ordinary Higgs field: one for leptons and two for quarks. Let us begin with lepton case.

The noncommutative Yukawa term for leptons reads~\cite{Calmet:2001na}
\begin{equation*}
\sum_{f_1 f_2}\idx\, Y_{f_1f_2}^{(lepton)}\,\overline{\hat{L}}^{(f_1)}_L\star\Phi_{lepton}\star\hat{e}^{(f_2)}_R.
\end{equation*}
Here, the noncommutative Higgs field, $\Phi_{lepton}$, is defined by the following hybrid Seiberg-Witten map
\begin{equation*}
\Phi_{lepton}(x)=\phi(x)+\Phi^{(1)}_{lepton}(x)+\Phi^{(2)}_{lepton}(x)+\Phi^{(3)}_{lepton}(x)+....,
\end{equation*}
where
\begin{equation*}
\begin{array}{l}
{\Phi^{(1)}_{lepton}(x)=\phantom{+}\int\frac{{\rm d}^4 p_{1}}{(2\pi)^4}\frac{{\rm d}^4 p_{2}}{(2\pi)^4}e^{-i(p_{1}+p_{2})x}\;\theta^{ij}p_{2j}\frac{e^{-i\frac{h}{2}(p_{1}\wedge p_{2})}-1}{p_{1}\wedge p_{2}}\;
[-\frac{1}{2}g^{'}\,a_i(p_1) \phi(p_2) + g\, b_i(p_1)\phi(p_2)]}\\[12pt]
{\phantom{\Phi^{(1)\,i_{L}}_{\phantom{(1)}\,i_{R}}(x)=}
+\int\frac{{\rm d}^4 p_{1}}{(2\pi)^4}\frac{{\rm d}^4 p_{2}}{(2\pi)^4}e^{-i(p_{1}+p_{2})x}\;\theta^{ij}p_{2j}\frac{e^{-i\frac{h}{2}(p_{2}\wedge p_{1})}-1}{p_{2}\wedge p_{1}}\;
 [g^{'}\,a_{i}(p1)\phi(p_2)]},
\end{array}
\end{equation*}
\begin{equation*}
\begin{array}{l}
{\Phi^{(2)}_{lepton}(x) =\int\prod_{i=1}^{3}\frac{{\rm d}^4 p_{i}}{(2\pi)^{4}}\;e^{-i(p_{1}+p_{2}+p_{3})x}\;}\\[12pt]
{
\Big\{\mathbb{M}^{(2,0)}[(\mu_{1},p_{1});(\mu_{2},p_{2});p_{3};h\theta]
\Big(\big(-\frac{1}{2}g^{'}a_{\mu_1}(p_1)\unit_2+g b_{\mu_1}(p_1)\big)
\big(-\frac{1}{2}g^{'}a_{\mu_2}(p_2)\unit_2+g b_{\mu_2}(p_2)\big)\Big)\phi(p_3)}\\[12pt]
{
+\mathbb{M}^{(1,1)}[(\mu_1,p_1);(\mu_2,p_2);p_3;h\theta]
\Big(\big(-\frac{1}{2}g^{'}a_{\mu_1}(p_1)\unit_2+g b_{\mu_1}(p_1)\big)g^{'}a_{\mu_2}(p_2)\unit_2\Big)\phi(p_3)}\\[12pt]
{+\mathbb{M}^{(0,2)}[(\mu_{1},p_{1});(\mu_{2},p_{2});p_{3};h\theta]\Big((g^{'})^2\, a_{\mu_1}(p_1)a_{\mu_2}(p_2)\unit_2\Big)\phi(p_3)}\Big\}
\end{array}
\end{equation*}
and
 \begin{equation*}
\begin{array}{l}
{\Phi^{(3)}_{lepton}(x) =\int\prod_{i=1}^{4}\frac{{\rm d}^4 p_{i}}{(2\pi)^{4}}\;e^{-i(p_{1}+p_{2}+p_{3}+p_{4})x}\;}\\[12pt]
{\Big\{\mathbb{M}^{(3,0)}[(\mu_{1},p_{1});(\mu_{2},p_{2});(\mu_{3},p_{3});p_{4};h\theta]}\\[12pt]
{\Big(\big(-\frac{1}{2}g^{'}a_{\mu_1}(p_1)\unit_2+g b_{\mu_1}(p_1)\big)
\big(-\frac{1}{2}g^{'}a_{\mu_2}(p_2)\unit_2+g b_{\mu_2}(p_2)\big)\big(-\frac{1}{2}g^{'}a_{\mu_3}(p_3)\unit_2+g b_{\mu_3}(p_3)\big)\Big)\phi(p_4)}\\[12pt]
{
+\mathbb{M}^{(2,1)}[(\mu_1,p_1);(\mu_2,p_2);(\mu_3,p_3);p_4;h\theta]}\\[12pt]
{\Big(\big(-\frac{1}{2}g^{'}a_{\mu_1}(p_1)\unit_2+g b_{\mu_1}(p_1)\big)
\big(-\frac{1}{2}g^{'}a_{\mu_2}(p_2)\unit_2+g b_{\mu_2}(p_2)\big)g^{'}a_{\mu_3}(p_3)\unit_2\Big)\phi(p_4)}\\[12pt]
{+\mathbb{M}^{(1,2)}[(\mu_1,p_1);(\mu_2,p_2);(\mu_3,p_3);p_4;h\theta]
\Big(\big(-\frac{1}{2}g^{'}a_{\mu_1}(p_1)\unit_2+g b_{\mu_1}(p_1)\big)
(g^{'})^2\,a_{\mu_2}(p_2)a_{\mu_3}(p_3)\unit_2\Big)\phi(p_4)}\\[12pt]
{+\mathbb{M}^{(0,3)}[(\mu_{1},p_{1});(\mu_{2},p_{2});(\mu_{3},p_{3});p_{4};h\theta] (g^{'})^3\, a_{\mu_1}(p_1) a_{\mu_2}(p_2) a_{\mu_3}(p_3)\unit_2\phi(p_4)
\Big\}.}
\end{array}
\end{equation*}

The noncommutative Yukawa term for the down-type quarks is~\cite{Calmet:2001na}
 \begin{equation}
\sum_{f_1 f_2}\idx\, Y_{f_1f_2}^{(down)}\,\overline{\hat{Q}}^{(f_1)}_{s_1\,L}\star\Phi^{\phantom{down}\,s_1}_{down\,s_2}\star\hat{d}^{(f_2)\,s_2}_{R}.
\label{downtype}
\end{equation}
In the previous expression, the indices $s_1$ and $s_2$ run from $1$ to $3$, since the ordinary quarks are in the fundamental representation of $SU(3)$. The noncommutative Higgs field, $\Phi^{\phantom{down}\,s_1}_{down\,s_2}(x)$, in (\ref{downtype}) is defined  by the hybrid Seiberg-Witten map, with expansion
\begin{equation*}
\Phi^{\phantom{down}\,s_1}_{down\,s_2}(x)=\phi(x)\,\delta^{s_1}_{s_2}+\Phi^{(1)\phantom{\!wn}\,s_1}_{down\,s_2}(x)+\Phi^{(2)\phantom{\!wn}\,s_1}_{down\,s_2}(x)
+\Phi^{(3)\phantom{\!wn}\,s_1}_{down\,s_2}(x)+....,
\end{equation*}
that is obtained by setting $z_d=1/3$ in the following expressions:
\begin{equation}
\begin{array}{l}
{\Phi^{(1)\phantom{\!wn}\,s_1}_{down\,s_2}(x)=\phantom{+}\int\frac{{\rm d}^4 p_{1}}{(2\pi)^4}\frac{{\rm d}^4 p_{2}}{(2\pi)^4}e^{-i(p_{1}+p_{2})x}\;\theta^{ij}p_{2j}\frac{e^{-i\frac{h}{2}(p_{1}\wedge p_{2})}-1}{p_{1}\wedge p_{2}}}\\[12pt]
{\phantom{\int\frac{{\rm d}^4 p_{1}}{(2\pi)^4}\frac{{\rm d}^4 p_{2}}{(2\pi)^4}e^{-i(p_{1}+p_{2})x}}
[\frac{1}{6}g^{'}\,a_i(p_1) \phi(p_2)\,\delta^{s_1}_{s_2} + g\, b_i(p_1)\phi(p_2)\,\delta^{s_1}_{s_2}+g_s\,G^{\phantom{i}\,s_1}_{i\,s_2}(p_1)\,\phi(p_2)]}\\[12pt]
{\phantom{\Phi^{(1)\,i_{L}}}
+\int\frac{{\rm d}^4 p_{1}}{(2\pi)^4}\frac{{\rm d}^4 p_{2}}{(2\pi)^4}e^{-i(p_{1}+p_{2})x}\;\theta^{ij}p_{2j}\frac{e^{-i\frac{h}{2}(p_{2}\wedge p_{1})}-1}{p_{2}\wedge p_{1}}\;
 [z_d\,g^{'}\,a_{i}(p_1)\phi(p_2)\,\delta^{s_1}_{s_2}-g_s\,G^{\phantom{i}\,s_1}_{i\,s_2}(p_1)\,\phi(p_2)]},
\end{array}
\label{downh1}
\end{equation}
\begin{equation}
\begin{array}{l}
{\Phi^{(2)\phantom{\!wn}\,s_1}_{down\,s_2}(x)
=\int\prod_{i=1}^{3}\frac{{\rm d}^4 p_{i}}{(2\pi)^{4}}\;e^{-i(p_{1}+p_{2}+p_{3})x}\;}\\[12pt]
{\Big\{\mathbb{M}^{(2,0)}[(\mu_{1},p_{1});(\mu_{2},p_{2});p_{3};h\theta]
\Big(\big(\frac{1}{6}g^{'}a_{\mu_1}(p_1)\delta^{s_1}_{s_3}\unit_2+g b_{\mu_1}(p_1)\delta^{s_1}_{s_3}+g_s G^{\phantom{\mu_1}\,s_1}_{\mu_1\,s_3}(p_1)\unit_2\big)}\\[12pt]
{\phantom{\Big\{\mathbb{M}^{(2,0)}[(\mu_{1},}
\big(\frac{1}{6}g^{'}a_{\mu_2}(p_2)\delta^{s_3}_{s_2}\unit_2+g b_{\mu_2}(p_2)\delta^{s_3}_{s_2}+g_s G^{\phantom{\mu_2}\,s_3}_{\mu_2\,s_2}(p_2)\unit_2\big)\Big)\phi(p_3)}\\[12pt]
{
+\mathbb{M}^{(1,1)}[(\mu_1,p_1);(\mu_2,p_2);p_3;h\theta]
\Big(\big(\frac{1}{6}g^{'}a_{\mu_1}(p_1)\delta^{s_1}_{s_3}\unit_2+g b_{\mu_1}(p_1)\delta^{s_1}_{s_3}+g_s G^{\phantom{\mu_1}\,s_1}_{\mu_1\,s_3}(p_1)\unit_2\big)}\\[12pt]
{\phantom{\Big\{\mathbb{M}^{(1,1)}[(\mu_{1},}
\big(z_dg^{'}a_{\mu_2}(p_2)\delta^{s_3}_{s_2}\unit_2-g_s G^{\phantom{\mu_2}\,s_3}_{\mu_2\,s_2}(p_2)\unit_2\big)\Big)\phi(p_3)}\\[12pt]
{+\mathbb{M}^{(0,2)}[(\mu_{1},p_{1});(\mu_{2},p_{2});p_{3};h\theta]
\Big(\big(z_dg^{'}a_{\mu_1}(p_1)\delta^{s_1}_{s_3}\unit_2-g_s G^{\phantom{\mu_1}\,s_1}_{\mu_1\,s_3}(p_1)\unit_2\big)}\\[12pt]
{\phantom{\Big\{\mathbb{M}^{(1,1)}[(\mu_{1},}
\big(z_dg^{'}a_{\mu_2}(p_2)\delta^{s_3}_{s_2}\unit_2-g_s G^{\phantom{\mu_2}\,s_3}_{\mu_2\,s_2}(p_2)\unit_2\big)\Big)\phi(p_3)\Big\}}
\end{array}
\label{downh2}
\end{equation}
and
\begin{equation}
\begin{array}{l}
{\Phi^{(3)\phantom{\!wn}\,s_1}_{down\,s_2}(x)=\int\prod_{i=1}^{4}\frac{{\rm d}^4 p_{i}}{(2\pi)^{4}}\;e^{-i(p_{1}+p_{2}+p_{3}+p_{4})x}\;}\\[12pt]
{\Big\{\mathbb{M}^{(3,0)}[(\mu_{1},p_{1});(\mu_{2},p_{2});(\mu_{3},p_{3});p_{4};h\theta]}\\[12pt]
{\Big(\big(\frac{1}{6}g^{'}a_{\mu_1}(p_1)\delta^{s_1}_{s_3}\unit_2+g b_{\mu_1}(p_1)\delta^{s_1}_{s_3}+g_s G^{\phantom{\mu_1}\,s_1}_{\mu_1\,s_3}(p_1)\unit_2\big)
\big(\frac{1}{6}g^{'}a_{\mu_2}(p_2)\delta^{s_3}_{s_4}\unit_2+g b_{\mu_2}(p_2)\delta^{s_3}_{s_4}+g_s G^{\phantom{\mu_2}\,s_3}_{\mu_1\,s_4}(p_2)\unit_2\big)}\\[12pt]
{\phantom{\Big(}
\big(\frac{1}{6}g^{'}a_{\mu_3}(p_3)\delta^{s_4}_{s_2}\unit_2+g b_{\mu_3}(p_3)\delta^{s_4}_{s_2}+g_s G^{\phantom{\mu_3}\,s_4}_{\mu_1\,s_2}(p_3)\unit_2\big)
\Big)\phi(p_4)}\\[12pt]
{+\mathbb{M}^{(2,1)}[(\mu_1,p_1);(\mu_2,p_2);(\mu_3,p_3);p_4;h\theta]
\Big(\big(\frac{1}{6}g^{'}a_{\mu_1}(p_1)\delta^{s_1}_{s_3}\unit_2+g b_{\mu_1}(p_1)\delta^{s_1}_{s_3}+g_s G^{\phantom{\mu_1}\,s_1}_{\mu_1\,s_3}(p_1)\unit_2\big)}\\[12pt]
{\phantom{+\mathbb{M}^{(2,1)}}
\big(\frac{1}{6}g^{'}a_{\mu_2}(p_2)\delta^{s_3}_{s_4}\unit_2+g b_{\mu_2}(p_2)\delta^{s_3}_{s_4}+g_s G^{\phantom{\mu_2}\,s_3}_{\mu_1\,s_4}(p_2)\unit_2\big)
\big(z_dg^{'}a_{\mu_3}(p_3)\delta^{s_4}_{s_2}\unit_2-g_s G^{\phantom{\mu_3}\,s_4}_{\mu_1\,s_2}(p_3)\unit_2\big)
\Big)\phi(p_4)}\\[12pt]
{+\mathbb{M}^{(1,2)}[(\mu_1,p_1);(\mu_2,p_2);(\mu_3,p_3);p_4;h\theta]
\Big(\big(\frac{1}{6}g^{'}a_{\mu_1}(p_1)\delta^{s_1}_{s_3}\unit_2+g b_{\mu_1}(p_1)\delta^{s_1}_{s_3}+g_s G^{\phantom{\mu_1}\,s_1}_{\mu_1\,s_3}(p_1)\unit_2\big)}\\[12pt]
{\phantom{+\mathbb{M}^{(1,2)}}
\big(z_dg^{'}a_{\mu_2}(p_2)\delta^{s_3}_{s_4}\unit_2-g_s G^{\phantom{\mu_2}\,s_3}_{\mu_1\,s_4}(p_2)\unit_2\big)
\big(z_dg^{'}a_{\mu_3}(p_3)\delta^{s_4}_{s_2}\unit_2-g_s G^{\phantom{\mu_3}\,s_4}_{\mu_1\,s_2}(p_3)\unit_2\big)
\Big)\phi(p_4)}\\[12pt]
{+\mathbb{M}^{(0,3)}[(\mu_{1},p_{1});(\mu_{2},p_{2});(\mu_{3},p_{3});p_{4};h\theta]
\Big(\big(z_dg^{'}a_{\mu_1}(p_1)\delta^{s_1}_{s_3}\unit_2-g_s G^{\phantom{\mu_1}\,s_1}_{\mu_1\,s_3}(p_1)\unit_2\big)}\\[12pt]
{\phantom{+\mathbb{M}^{(0,3)}}
\big(z_dg^{'}a_{\mu_2}(p_2)\delta^{s_3}_{s_4}\unit_2-g_s G^{\phantom{\mu_2}\,s_3}_{\mu_1\,s_4}(p_2)\big)
\big(z_dg^{'}a_{\mu_3}(p_3)\delta^{s_4}_{s_2}\unit_2-g_s G^{\phantom{\mu_3}\,s_4}_{\mu_1\,s_2}(p_3)\unit_2\big)
\Big)\phi(p_4)
\Big\}.}
\end{array}
\label{downh3}
\end{equation}

Finally, the noncommutative Yukawa term for the up-type  quarks~\cite{Calmet:2001na} reads
 \begin{equation*}
\sum_{f_1 f_2}\idx\, Y_{f_1f_2}^{(up)}\,\overline{\hat{Q}}^{(f_1)}_{s_1\,L}\star\Phi^{\phantom{up}\,s_1}_{up\,s_2}\star\hat{u}^{(f_2)\,s_2}_{R}.
\end{equation*}
The noncommutative Higgs field, $\Phi^{\phantom{up}\,s_1}_{up\,s_2}$, is a hybrid Seiberg-Witten map with an expansion in the number of gauge fields,
\begin{equation*}
\Phi^{\phantom{up}\,s_1}_{up\,s_2}(x)=i\tau_2\phi(x)\,\delta^{s_1}_{s_2}+\Phi^{(1)\,s_1}_{up\,s_2}(x)+\Phi^{(2)\,s_1}_{up\,s_2}(x)
+\Phi^{(3)\,s_1}_{up\,s_2}(x)+....,
\end{equation*}
whose terms $\Phi^{(1)\phantom{\!up}\,s_1}_{up\,s_2}(x)$, $\Phi^{(2)\phantom{\!up}\,s_1}_{up\,s_2}(x)$ and
$\Phi^{(3)\phantom{\!up}\,s_1}_{up\,s_2}(x)$ are obtained by setting $z_d=-2/3$ and replacing $\phi$ with $i\tau_2\phi$ in (\ref{downh1}),  (\ref{downh2}) and (\ref{downh3}), respectively.

To close this section we shall derive, following ref.~\cite{Aschieri:2012in}, a $\theta$-exact expression of a general Yukawa term of the form
\begin{equation*}
S_{Yukawa}[\theta^{\mu\nu}]=\idx\,\Psi\star\Phi\star\chi,
\end{equation*}
where $\Phi$ is a noncommutative scalar field defined by the hybrid Seiberg-Witten map in (\ref{hybridevol}) and $\Psi$ and $\chi$ are noncommutative spinor fields defined by the following equations
\begin{equation}
\begin{array}{l}
{{\displaystyle \frac{d\Psi}{dh}}= \frac{1}{2}\,\theta^{ij}\,\partial_j\Psi\star_{h} A_i- \frac{i}{4}\,\theta^{ij}\,\Psi\star_{h} A_j\star_{h} A_i}\\[8pt]
{{\displaystyle \frac{d\xi}{dh}}=\frac{1}{2}\,\theta^{ij}\,B_i\star_{h}\partial_j\chi+\frac{i}{4}\,\theta^{ij}\,B_i\star_{h} B_j\star_{h}\chi}.
\end{array}
\label{fermyuk}
\end{equation}
Notice that $\Psi$ and $\chi$ transforms under noncommutative BRS transformations as follows
\begin{equation*}
s_{NC}\Psi=i\Phi\star_h\Lambda,\quad s_{NC}\chi=-i\Omega\star_h\chi,
\end{equation*}
so that $S_{Yukawa}$ is BRS invariant.

Now replacing $\theta^{\mu\nu}$ with $h\theta^{\mu\nu}$ in $S_{Yukawa}[\theta^{\mu\nu}]$ and using  (\ref{hybridevol}) and (\ref{fermyuk}), one obtains after some algebra
\begin{equation*}
\begin{array}{l}
{{\displaystyle \frac{d S_{Yukawa}[h\theta]}{dh}}=
\idx\;\big(-\frac{i}{2}\theta^{ij}\,D_i\Psi[h]\star_h\Phi[h]\star_h D_j\chi[h] +\frac{\theta^{ij}}{4}\,\Psi[h]\star_h A_{ij}[h]\star_h\Phi[h]\star_h\chi[h]}\\[8pt]
{\phantom{{\displaystyle \frac{d S_{Yukawa}[h\theta]}{dh}}=\idx\;\big(} +\frac{\theta^{ij}}{4}\,\Psi[h]\star_h B_{ij}[h]\star_h\Phi[h]\star_h\chi[h]\big).}
\end{array}
\end{equation*}
$A_{\mu\nu}$ and $B_{\mu\nu}$ are the field strengths of $A_{\mu}$ and $B_{\mu}$, respectively.
Integrating both sides of the previous equation with respect to $h$ one gets
\begin{equation}
\begin{array}{l}
{S_{Yukawa}[\theta^{\mu\nu}]=S_{Yukawa}[0]+S_{nccorrection}[\theta^{\mu\nu}],}\\[8pt]
{S_{nccorrection}[\theta^{\mu\nu}]=\idx\,\int_{0}^{1}dh\,\big(-\frac{i}{2}\theta^{ij}\,D_i\Psi[h]\star_h\Phi[h]\star_h D_j\chi[h] +\frac{\theta^{ij}}{4}\Psi[h]\star_h A_{ij}[h]\star_h\Phi[h]\star_h\chi[h]}\\[8pt]
{\phantom{S_{nccorrection}[\theta^{\mu\nu}]=\idx\,\int_{0}^{1}dh\,\big(}
+\frac{\theta^{ij}}{4}\Psi[h]\star_h B_{ij}[h]\star_h\Phi[h]\star_h\chi[h]\big)},
\end{array}
\label{ncyuka}
\end{equation}
where $S_{Yukawa}[0]$ is the Yukawa term on ordinary space-time. The previous expression is a $\theta$-exact closed expression for $S_{Yukawa}[\theta^{\mu\nu}]$ in terms of the $\theta$-exact Seiberg-Witten maps where the full $\theta$-exact noncommutative correction to the ordinary Yukawa term has
been isolated and can be used to iteratively compute such correction as powers in the number of ordinary gauge fields. Notice that what  (\ref{ncyuka}) shows is that the noncommutative Yukawa  correction, $S_{nccorrection}[\theta^{\mu\nu}]$, has a beautiful expression in terms of the noncommutative differential-geometric objects --namely, the gauge curvatures and the covariant derivatives-- and is thus explicitly gauge invariant.
The particularization of the previous expression to the Noncommutative Standard model is straightforward.

\section{Final comments and outlook}

In this paper we have shown how the antifield formalism can be successfully used to derive an ``evolution" equation for the hybrid Seiberg-Witten map for arbitrary compact gauge groups in arbitrary faithful matrix representations, thus implying that noncommutative gauge theories with hybrid Seiberg-Witten map are consistent deformations of ordinary gauge theories in the sense of~\cite{Barnich:2001mc}. We have also shown that this ``evolution" equation can be solved recursively in a $\theta$-exact way, thus providing a tool to systematically construct hybrid Seiberg-Witten maps which will give rise to UV/IR mixing effects. We have computed the expansion of a general $\theta$-exact hybrid Seiberg-Witten up to order three in the number of ordinary gauge fields. Finally, we have worked out explicitly, up to three ordinary gauge fields, the three $\theta$-exact hybrid Seiberg-Witten map that are needed to formulate the Yukawa terms of the noncommutative Standard Model. We also derive the general expression of the $\theta$-exact noncommutative corrections to a general ordinary Yukawa term in terms of noncommutative field strengths and covariant derivatives. Furnished with formulae presented in this paper --along with the results in Ref.~\cite{Martin:2012aw}-- a systematic study of the occurrence of noncommutative effects --UV/IR mixing phenomena, in particular-- on  the physics  of the Higgs particle and  other particles of the Standard Model can be launched. Besides, the  equivalence, at the quantum level, of supersymmetric noncommutative $U(n)$ gauge theories formulated in terms of noncommutative fields and the same classical theories formulated, by means of the Seiberg-Witten map, in terms of ordinary fields can be systematically analyzed for  matter in the fundamental, anti-fundamental and bi-fundamnetal representations.

\newpage

\section{Appendix A}

Let $\Phi^{i_L}_{i_R}$ be a boson field. Let
\begin{equation*}
A_{\mu\nu}=\partial_{\mu}A_{\nu}-\partial_{\nu}A_{\mu}+i[A_{\mu},A_{\nu}]_{\star},\quad B_{\mu\nu}=\partial_{\mu}B_{\nu}-\partial_{\nu}B_{\mu}+i[B_{\mu},B_{\nu}]_{\star},\quad
D_\mu\Phi=\partial_{\mu}\Phi+i A_\mu\star\Phi-i\Phi\star B_\mu.
\end{equation*}
Then, a standard  functional that is invariant under the BRST transformations in (\ref{ABRST}), (\ref{BBRST}) and (\ref{HBRST}) reads
\begin{equation*}
\hat{S}_{0}=-\frac{1}{4g^2_A}\idx\,Tr A_{\mu\nu}\star A^{\mu\nu}-\frac{1}{4g^2_B}\idx\,Tr B_{\mu\nu}\star B^{\mu\nu}+\idx\, (D_\mu\bar{\Phi})^{i_R}_{i_L}\star (D_\mu\Phi)^{i_L}_{i_R} - V[\Phi^{i_L}_{i_R}, \bar{\Phi}^{i_R}_{i_L}],
\end{equation*}
where
\begin{equation*}
 V[\Phi^{i_L}_{i_R}, \bar{\Phi}^{i_R}_{i_L}]=\pm M^2\idx\;  \bar{\Phi}^{i_R}_{i_L}\star\Phi^{i_L}_{i_R}+\lambda \idx\,  \bar{\Phi}^{i_R}_{j_L}\star\,\Phi^{j_L}_ {i^{'}_R}\star\,\bar{\Phi}^{i^{'}_R}_{j^{'}_L}\,\star\Phi^{j^{'}_L}_{i_R}.
\end{equation*}
In the equations above repeated indices indicates sum over all their values.

\section{Appendix B}

Here we shall give some details of the computation of $\hat{\cal A}[F^M,F^{*}_M;h\theta]$ in (\ref{defantibit}) and (\ref{antibit}). We shall focus on the contributions that are linear in the antifields $\Phi^{*\,i_R}_{\phantom{*}i_L}$

The antibracket $(\hat{\cal J},\hat{S}_{Antifields})$ is defined, in full detail, as follows
\begin{equation}
\begin{array}{l}
{(\hat{\cal J},\hat{S}_{Antifields})=}\\[8pt]
{\phantom{(\hat{\cal J},}
\idx\;\Big[\phantom{+}\frac{\partial_r \hat{\cal J}}{\partial A^a_\mu}\frac{\partial_l \hat{S}_{Antifields}}{\partial A^{*\,\mu}_a}-\frac{\partial_r \hat{\cal J}}{\partial A^{*\,\mu}_a }\frac{\partial_l \hat{S}_{Antifields}}{\partial A^{a}_{\mu}}+
\frac{\partial_r \hat{\cal J}}{\partial B^a_\mu}\frac{\partial_l \hat{S}_{Antifields}}{\partial B^{*\,\mu}_a}-\frac{\partial_r \hat{\cal J}}{\partial B^{*\,\mu}_a }\frac{\partial_l \hat{S}_{Antifields}}{\partial B^{a}_{\mu}}}\\[8pt]
{\phantom{(\hat{\cal J},\idx\;\Big[}
+\frac{\partial_r \hat{\cal J}}{\partial \Lambda^a}\frac{\partial_l \hat{S}_{Antifields}}{\partial \Lambda^{*}_a}-\frac{\partial_r \hat{\cal J}}{\partial \Lambda^{*}_a }\frac{\partial_l \hat{S}_{Antifields}}{\partial \Lambda^{a}}+
\frac{\partial_r \hat{\cal J}}{\partial \Omega^a}\frac{\partial_l \hat{S}_{Antifields}}{\partial \Omega^{*}_a}-\frac{\partial_r \hat{\cal J}}{\partial \Omega^{*}_a }\frac{\partial_l \hat{S}_{Antifields}}{\partial \Omega^{a}}
}\\[8pt]
{\phantom{(\hat{\cal J},\idx\;\Big[}
+\frac{\partial_r \hat{\cal J}}{\partial \Phi^{i_L}_{i_R}}\frac{\partial_l \hat{S}_{Antifields}}{\partial \Phi^{*\,i_R}_{\phantom{*}i_L}}-\frac{\partial_r \hat{\cal J}}{\partial \Phi^{*\,i_R}_{\phantom{*}i_L} }\frac{\partial_l \hat{S}_{Antifields}}{\partial \Phi^{i_L}_{i_R} }+
\frac{\partial_r \hat{\cal J}}{\partial \bar{\Phi}^{i_R}_{i_L}}\frac{\partial_l \hat{S}_{Antifields}}{\partial\bar{\Phi}^{*\,i_L}_{\phantom{*}i_R}}-\frac{\partial_r \hat{\cal J}}{\partial\bar{\Phi}^{*\,i_L}_{\phantom{*}i_R} }\frac{\partial_l \hat{S}_{Antifields}}
{\partial\bar{\Phi}^{i_R}_{i_L}}\big].
}
\end{array}
\label{expandantib}
\end{equation}

Let us display the contributions to the previous equation which are linear in  $\Phi^{*\,i_R}_{\phantom{*}i_L}$:
\begin{equation*}
\begin{array}{l}
{\idx\;\frac{\partial_r \hat{\cal J}}{\partial A^a_\mu}\frac{\partial_l \hat{S}_{Antifields}}{\partial A^{*\,\mu}_a}=}\\[8pt]
{
-\idx\;\Phi^{*\,i_R}_{\phantom{*}i_L}\big(\frac{\theta^{ij}}{2}D_i\Lambda\star_{h}\partial_j\Phi+i\frac{\theta^{ij}}{4}D_i\Lambda\star_{h}A_j\star_{h}\Phi
+i\frac{\theta^{ij}}{4}A_i\star_{h} D_j\Lambda\star_{h}\star_{h}\Phi-i\frac{\theta^{ij}}{2} D_i\Lambda\star_{h}\Phi\star_h B_j\big)^{i_L}_{i_R},}\\[8pt]
{\idx\;\frac{\partial_r \hat{\cal J}}{\partial B^a_\mu}\frac{\partial_l \hat{S}_{Antifields}}{\partial B^{*\,\mu}_a}=}\\[8pt]
{-\idx\;\Phi^{*\,i_R}_{\phantom{*}i_L}\big(\frac{\theta^{ij}}{2}\partial_j\Phi\star_h D_i\Omega-i\frac{\theta^{ij}}{4}\Phi\star_{h} D_j\Omega\star_h B_i
-i\frac{\theta^{ij}}{4}\Phi\star_h B_j\star_{h} D_i\Omega-i\frac{\theta^{ij}}{2} A_i\star_{h}\Phi\star_h D_j\Omega\big)^{i_L}_{i_R},}\\[8pt]
{\idx\;\frac{\partial_r \hat{\cal J}}{\partial \Lambda^a}\frac{\partial_l \hat{S}_{Antifields}}{\partial \Lambda^{*}_a}=
-\idx\;\Phi^{*\,i_R}_{\phantom{*}i_L}i\frac{\theta^{ij}}{4}\big(\partial_i\Lambda\star_h A_j\star_h\Phi+ A_j\star_h\partial_i\Lambda\star_h\Phi\big)^{i_L}_{i_R},}\\[8pt]
{\idx\;\frac{\partial_r \hat{\cal J}}{\partial \Omega^a}\frac{\partial_l \hat{S}_{Antifields}}{\partial \Omega^{*}_a}=
\idx\;\Phi^{*\,i_R}_{\phantom{*}i_L}i\frac{\theta^{ij}}{4}\big(\Phi\star_h\partial_i\Omega B_j+\Phi\star_h B_j\star_h\partial_i\Omega\big)^{i_L}_{i_R},}\\[8pt]
\idx\;{\frac{\partial_r \hat{\cal J}}{\partial \Phi^{i_L}_{i_R}}\frac{\partial_l \hat{S}_{Antifields}}{\partial \Phi^{*\,i_R}_{\phantom{*}i_L}}=}\\[8pt]
{-\idx\;\Big[\Phi^{*\,i_R}_{\phantom{*}i_L}\big(\frac{\theta^{ij}}{2}A_i\star_h[-i\Lambda\star_h\Phi+i\Phi\star_h\Omega]
+i\frac{\theta^{ij}}{4} A_i\star_h A_j\star_h [-i\Lambda\star_h\Phi+i\Phi\star_h\Omega]}\\[8pt]
{\phantom{-\idx\;\Big[\Phi^{*\,i_R}_{\phantom{*}i_L}\big(\frac{\theta^ij}{4}    }
+
\frac{\theta^{ij}}{2}\partial_j[-i\Lambda\star_h\Phi+i\Phi\star_h\Omega]\star_h B_i
-i\frac{\theta^{ij}}{4}[-i\Lambda\star_h\Phi+i\Phi\star_h\Omega]\star_h B_j\star_h B_i}\\[8pt]

{\phantom{-\idx\;\Big[\Phi^{*\,i_R}_{\phantom{*}i_L}\big(\frac{\theta^ij}{2} }
-\frac{\theta^{ij}}{2}A_i\star_h[-i\Lambda\star_h\Phi+i\Phi\star_h\Omega]\star_h B_j\big)^{i_L}_{i_R}\Big],
}\\[8pt]
\idx\;{\frac{\partial_r \hat{\cal J}}{\partial\Phi^{*\,i_R}_{\phantom{*}i_L} }\frac{\partial_l \hat{S}_{Antifields}}{\partial\Phi^{i_L}_{i_R} }=\idx\,\Phi^{*\,i_R}_{\phantom{*}i_L}\Big[}\\[8pt]
{\big(i\Lambda\star_h[\frac{\theta^{ij}}{2}A_i\star_h\partial_j\Phi
+i\frac{\theta^{ij}}{4} A_i\star_h A_j\star_h \Phi+
\frac{\theta^{ij}}{2}\partial_j\Phi\star_h B_i
-i\frac{\theta^{ij}}{4}\Phi\star_h B_j\star_h B_i
-\frac{\theta^{ij}}{2}A_i\star_h\Phi\star_h B_j]\big)^{i_L}_{i_R}}\\[8pt]
{+\big([\frac{\theta^{ij}}{2}A_i\star_h\partial_j\Phi
+i\frac{\theta^{ij}}{4} A_i\star_h A_j\star_h \Phi+
\frac{\theta^{ij}}{2}\partial_j\Phi\star_h B_i
-i\frac{\theta^{ij}}{4}\Phi\star_h B_j\star_h B_i
-\frac{\theta^{ij}}{2}A_i\star_h\Phi\star_h B_j]\star_h\Omega\big)^{i_L}_{i_R}\Big].
}
\end{array}
\end{equation*}

The substitution of the previous results in (\ref{expandantib}) and some lengthy algebra yields that the contribution to $\hat{\cal A}[F^M,F^{*}_M;h\theta]$ that is linear in  $\Phi^{*\,i_R}_{\phantom{*}i_L}$ reads
\begin{equation*}
\idx\; \Phi^{*\,i_R}_{\phantom{*}i_L}\frac{\theta^{ij}}{2}\big(\partial_i\Lambda\star_h\partial_j\Phi-\partial_i\Phi\star_h\partial_j\Omega\big)^{i_L}_{i_R},
\end{equation*}
which matches the appropriate summands in the RHS of (\ref{antibit}). All the remaining  summands in the RHS of (\ref{antibit}) are obtained by carrying out similar algebraic computations.

\section{Appendix C}

In this Appendix we give the value $\mathbb{M}^{(3,0)}[(\mu_{1},p_{1});(\mu_{2},p_{2});(\mu_{3},p_{3});p_{4};h\theta]$,\\
$\mathbb{M}^{(3,0)}[(\mu_{1},p_{1});(\mu_{2},p_{2});(\mu_{3},p_{3});p_{4};h\theta]$, $\mathbb{M}^{(2,1)}[(\mu_{1},p_{1});(\mu_{2},p_{2});(\mu_{3},p_{3});p_{4};h\theta]$\\ and $\mathbb{M}^{(1,2)}[(\mu_{1},p_{1});(\mu_{2},p_{2});(\mu_{3},p_{3});p_{4};h\theta]$ in (\ref{3order}). Let us begin with some definitions

\begin{equation*}
\begin{array}{l}
{\mathbb{P}^{(3)}_{m}[(p_1,\mu_1),(p_2,\mu_2),(p_3,\mu_3);\theta]=\frac{1}{4}\,\theta^{ij}\theta^{kl}\Big\{
[4(p_{3 l}\,\delta_k^{\mu_2}\delta_i^{\mu_3}\!+\!p_{2 l}\,\delta_i^{\mu_2}\delta_k^{\mu_3})-2(p_3\!-\!p_2)_i\,\delta_k^{\mu_2} \delta_l^{\mu_3}]p_{1j}\,\delta_m^{\mu_1}}\\[4pt]
{\phantom{\mathbb{P}^{(3)}_{m}[(p_1,\mu_1),(p_2,\mu_2),(p_3,\mu_3)}
+[4(p_{3 l}\,\delta_k^{\mu_2}\delta_m^{\mu_3}+p_{2 l}\,\delta_m^{\mu_2}\delta_k^{\mu_3})-2(p_3\!-\!p_2)_{m}\,\delta_k^{\mu_2} \delta_l^{\mu_3}]\,(p_2\!+\!p_3)_j\,\delta_i^{\mu_1}}\\[4pt]
{\phantom{\mathbb{P}^{(3)}_{m}[(p_1,\mu_1),(p_2,\mu_2),(p_3,\mu_3)}
-[2(p_{3 l}\,\delta_k^{\mu_2}\delta_i^{\mu_3}+p_{2 l}\,\delta_i^{\mu_2}\delta_k^{\mu_3})-(p_3\!-\!p_2)_{i}\,\delta_k^{\mu_2} \delta_l^{\mu_3}]\,p_{1 m}\,\delta_{j}^{\mu_1}}\\[4pt]
{\phantom{\mathbb{P}^{(3)}_{m}[(p_1,\mu_1),(p_2,\mu_2),(p_3,\mu_3)}
-[2(p_{3 l}\,\delta_k^{\mu_2}\delta_j^{\mu_3}+p_{2 l}\,\delta_j^{\mu_2}\delta_k^{\mu_3})-(p_3\!-\!p_2)_{j}\,\delta_k^{\mu_2} \delta_l^{\mu_3}](p_2\!+\!p_3)_m\,\delta_{i}^{\mu_1}\Big\},}\\[8pt]
{\mathbb{Q}^{(3)}_{m}[\mu_1,\mu_2,\mu_3;\theta]=-\frac{1}{2}\,\theta^{ij}(\delta_i^{\mu_1}\delta_j^{\mu_2}\delta_m^{\mu_3}-\delta_i^{\mu_1}\delta_m^{\mu_2}\delta_j^{\mu_3}).}
\end{array}
\end{equation*}

\begin{equation*}
\begin{array}{l}
{\Sigma(p_1,p_2,p_3,p_4,\theta)=\sum_{i<j}\,p_1\wedge p_j=(p_1+p_2+p_3)\wedge p_4+p_2\wedge p_3 + p_1\wedge (p_2+p_3),}\\[10pt]
{\Theta(p_1,p_2,p_3,p_4,\theta)=(p_1+p_2+p_3)\wedge p_4+p_2\wedge p_3 - p_1\wedge (p_2+p_3),}\\[10pt]
{\mathbb{L}_1(p_1,p_2,p_3,p_4;h,\theta)=}\\[4pt]
{\quad\displaystyle{
\frac{1}{p_1\wedge(p_2+p_3)+p_2\wedge p_3}\Big[
\frac{e^{-i\frac{h}{2}\Sigma(p_1,p_2,p_3,p_4,\theta)}-1}{\Sigma(p_1,p_2,p_3,p_4,\theta)}-\frac{e^{-i\frac{h}{2}(p_1+p_2+p_3)\wedge p_4}-1}{(p_1+p_2+p_3)\wedge p_4}\Big],}}\\[18pt]
{\mathbb{L}_2(p_1,p_2,p_3,p_4;h,\theta)=}\\[4pt]
{\quad\displaystyle{\frac{1}{-p_1\wedge(p_2+p_3)+p_2\wedge p_3}\Big[
\frac{e^{-i\frac{h}{2}\Theta(p_1,p_2,p_3,p_4,\theta)}-1}{\Theta(p_1,p_2,p_3,p_4,\theta)}-\frac{e^{-i\frac{h}{2}(p_1+p_2+p_3)\wedge p_4}-1}{(p_1+p_2+p_3)\wedge p_4}\Big],}}\\[18pt]
{\mathbb{K}_1(p_1,p_2,p_3,p_4;h,\theta)=\displaystyle{
\frac{1}{p_2\wedge p_3}\Big\{\;\mathbb{L}_1(p_1,p_2,p_3;h,\theta)}}\\[8pt]
{\quad\displaystyle{-\frac{1}{p_1\wedge (p_2+p_3)}
\Big[\frac{e^{-i\frac{h}{2}[(p_1+p_2+p_3)\wedge p_4+p_1\wedge(p_2+p_3)]}-1}{(p_1+p_2+p_3)\wedge p_4+p_1\wedge(p_2+p_3)}-\frac{e^{-i\frac{h}{2}(p_1+p_2+p_3)\wedge p_4}-1}{(p_1+p_2+p_3)\wedge p_4}\Big]\Big\}},}\\[18pt]
{\mathbb{K}_2(p_1,p_2,p_3,p_4;h,\theta)=
\displaystyle{\frac{1}{p_2\wedge p_3}\Big\{\,\mathbb{L}_2(p_1,p_2,p_3;h,\theta) }}\\[8pt]
{\quad\displaystyle{+\frac{1}{p_1\wedge (p_2+p_3)}
\Big[\frac{e^{-i\frac{h}{2}[(p_1+p_2+p_3)\wedge p_4-p_1\wedge(p_2+p_3)]}-1}{(p_1+p_2+p_3)\wedge p_4-p_1\wedge(p_2+p_3)}-\frac{e^{-i\frac{h}{2}(p_1+p_2+p_3)\wedge p_4}-1}{(p_1+p_2+p_3)\wedge p_4}\Big]\Big\}},}\\[18pt]
{\mathbb{K}_3(p_1,p_2,p_3,p_4;h,\theta)=}\\[4pt]
{\quad\displaystyle{\frac{1}{(p_1\wedge p_2)(p_3\wedge p_4)} \Big[
\frac{e^{-i\frac{h}{2}\Sigma(p_1,p_2,p_3,p_4,\theta)}-1}{\Sigma(p_1,p_2,p_3,p_4,\theta)}-
\frac{e^{-i\frac{h}{2}[p_1\wedge p_2+(p_1+p_2)\wedge(p_3+ p_4)]}-1}{p_1\wedge p_2+(p_1+p_2)\wedge(p_3+ p_4)}}}\\[8pt]
{\quad\quad\quad\quad\quad\quad\quad\quad\quad\quad\displaystyle{-\frac{e^{-i\frac{h}{2}[p_3\wedge p_4+(p_1+p_2)\wedge(p_3+ p_4)]}-1}{p_3\wedge p_4+(p_1+p_2)\wedge(p_3+ p_4)}+
\frac{e^{-i\frac{h}{2} (p_1+p_2)\wedge(p_3+ p_4)}-1}{(p_1+p_2)\wedge(p_3+ p_4)}
\Big],}}\\[18pt]
\end{array}
\end{equation*}

\begin{equation*}
\begin{array}{l}
{\mathbb{K}_4(p_1,p_2,p_3, p_4;h,\theta)=}\\[4pt]
{\quad\displaystyle{
\frac{1}{p_3\wedge p_4}\Big\{
\frac{1}{p_2\wedge p_3+(p_2+p_3)\wedge p_4}\Big[
\frac{e^{-i\frac{h}{2}\Sigma(p_1,p_2,p_3,p_4,\theta)}-1}{\Sigma(p_1,p_2,p_3,p_4,\theta)}-\frac{e^{-i\frac{h}{2}p_1\wedge( p_2+p_3+ p_4)}-1}{p_1\wedge (p_2+p_3 + p_4)}\Big]}}\\[10pt]
{\quad\displaystyle{-\frac{1}{p_2\wedge (p_3+p_4)}
\Big[\frac{e^{-i\frac{h}{2}[p_1\wedge (p_2+p_3 + p_4)+p_2\wedge(p_3+p_4)]}-1}{p_1\wedge (p_2+p_3 + p_4)+p_2\wedge(p_3+p_4)}-\frac{e^{-i\frac{h}{2}p_1\wedge ( p_2+p_3 + p_4)}-1}{p_1\wedge ( p_2+p_3 + p_4)}\Big]\Big\}},}\\[18pt]
{\mathbb{K}_5(p_1,p_2,p_3, p_4;h,\theta)=}\\[4pt]
{\quad\displaystyle{
\frac{1}{p_2\wedge p_3}\Big\{
\frac{1}{p_2\wedge p_3+(p_2+p_3)\wedge p_4}\Big[
\frac{e^{-i\frac{h}{2}\Sigma(p_1,p_2,p_3,p_4,\theta)}-1}{\Sigma(p_1,p_2,p_3,p_4,\theta)}-\frac{e^{-i\frac{h}{2}p_1\wedge( p_2+p_3+ p_4)}-1}{p_1\wedge (p_2+p_3 + p_4)}\Big]}}\\[10pt]
{\quad\displaystyle{-\frac{1}{(p_2+p_3)\wedge p_4}
\Big[\frac{e^{-i\frac{h}{2}[p_1\wedge (p_2+p_3 + p_4)+(p_2+p_3)\wedge p_4]}-1}{p_1\wedge (p_2+p_3 + p_4)+(p_2+p_3)\wedge p_4}-\frac{e^{-i\frac{h}{2}p_1\wedge ( p_2+p_3 + p_4)}-1}{p_1\wedge ( p_2+p_3 + p_4)}\Big]\Big\}},}\\[18pt]
{\mathbb{K}_6(p_1,p_2,p_3, p_4;h,\theta)=
\displaystyle{\frac{1}{p_2\wedge p_3+(p_2+p_3)\wedge p_4}\Big[
\frac{e^{-i\frac{h}{2}\Sigma(p_1,p_2,p_3,p_4,\theta)}-1}{\Sigma(p_1,p_2,p_3,p_4,\theta)}-\frac{e^{-i\frac{h}{2}p_1\wedge( p_2+p_3+ p_4)}-1}{p_1\wedge (p_2+p_3 + p_4)}}\Big]}\\[18pt]
{\mathbb{K}_7(p_1,p_2,p_3,p_4;h,\theta)=\displaystyle{
\frac{1}{p_1\wedge p_2}\Big[
\frac{e^{-i\frac{h}{2}\Sigma(p_1,p_2,p_3,p_4,\theta)}-1}{\Sigma(p_1,p_2,p_3,p_4,\theta)}-\frac{e^{-i\frac{h}{2}[(p_1+p_2)\wedge (p_3+p_4)+p_3\wedge p_4]}-1}{(p_1+p_2)\wedge(p_3+p_4)+p_3\wedge p_4}\Big],}}\\[18pt]
{\mathbb{K}_8(p_1,p_2,p_3,p_4;h,\theta)=\displaystyle{
\frac{1}{p_2\wedge p_3}\Big[
\frac{e^{-i\frac{h}{2}\Sigma(p_1,p_2,p_3,p_4,\theta)}-1}{\Sigma(p_1,p_2,p_3,p_4,\theta)}-\frac{e^{-i\frac{h}{2}[p_1\wedge (p_2+p_3+p_4)+(p_2+p_3)\wedge p_4]}-1}{p_1\wedge(p_2+p_3+p_4)+(p_2+p_3)\wedge p_4}\Big],}}\\[18pt]
{\mathbb{K}_9(p_1,p_2,p_3,p_4;h,\theta)=\displaystyle{
\frac{1}{p_3\wedge p_4}\Big[
\frac{e^{-i\frac{h}{2}\Sigma(p_1,p_2,p_3,p_4,\theta)}-1}{\Sigma(p_1,p_2,p_3,p_4,\theta)}-\frac{e^{-i\frac{h}{2}[p_1\wedge (p_2+p_3+p_4)+p_2\wedge (p_3 + p_4)]}-1}{p_1\wedge(p_2+p_3+p_4)+ p_2\wedge( p_3 + p_4)}\Big].}}
\end{array}
\end{equation*}

Then

\begin{equation*}
\begin{array}{l}
{\mathbb{M}^{(3,0)}[(\mu_{1},p_{1});(\mu_{2},p_{2});(\mu_{3},p_{3});p_{4};h\theta]=\theta^{mn}\,p_{4n}}\\[8pt]
{\Big[\mathbb{P}^{(3)}_{m}[(p_1,\mu_1),(p_2,\mu_2),(p_3,\mu_3);\theta]\,
\mathbb{K}_1(p_1,p_2,p_3,p_4;h,\theta)
+\mathbb{Q}^{(3)}_{m}[\mu_1,\mu_2,\mu_3;\theta]\,\mathbb{L}_1(p_1,p_2,p_3,p_4;h,\theta)}\\[8pt]
{+\mathbb{P}^{(3)}_{m}[(p_3,\mu_3),(p_1,\mu_1),(p_2,\mu_2);\theta]\,
\mathbb{K}_2(p_3,p_1,p_2,p_4;h,\theta)
+\mathbb{Q}^{(3)}_{m}[\mu_3,\mu_1,\mu_2;\theta]\,
\mathbb{L}_2(p_3,p_1,p_2,p_4;h,\theta)\Big]}\\[8pt]
{+\theta^{ij}\theta^{mn}\theta^{kl}\;
\Big[ }\\[8pt]
{\quad\frac{1}{2}\,(p_3\!+\!p_4)_j
[2(p_{2 l}\,\delta_k^{\mu_1}\delta_i^{\mu_2}+p_{1 l}\,\delta_k^{\mu_2}\delta_i^{\mu_1})-(p_2\!-\!p_1)_{i}\,\delta_k^{\mu_1} \delta_l^{\mu_2}]\,\delta_{m}^{\mu_3}\,p_{4n}
\;\mathbb{K}_3(p_1,p_2,p_3,p_4;h,\theta) }\\[8pt]
{\quad\quad\quad\quad\quad +\delta_i^{\mu_1}\delta_{m}^{\mu_2}\delta_k^{\mu_3}\,(p_2\!+\!p_3\!+\!p_4)_j\,(p_3\!+\!p_4)_n\,p_{4l}\,
\;\mathbb{K}_4(p_1,p_2,p_3,p_4;h,\theta)}\\[8pt]
{\;+\frac{1}{2}\,\delta_{i}^{\mu_1}(p_2\!+\!p_3\!+\!p_4)_{j} p_{4n}\,
[2(p_{3 l}\,\delta_k^{\mu_2}\delta_m^{\mu_3}+p_{2 l}\,\delta_m^{\mu_2}\delta_k^{\mu_3})-(p_3\!-\!p_2)_{m}\,\delta_k^{\mu_2} \delta_l^{\mu_3}]
\;\mathbb{K}_5(p_1,p_2,p_3,p_4;h,\theta)\Big] }\\[8pt]
{\quad\quad\quad\quad\quad-\frac{1}{2}\,\theta^{ij}\theta^{kl}\,\delta_i^{\mu_1}\delta_{k}^{\mu_2}\delta_l^{\mu_3}\,(p_2\!+\!p_3\!+\!p_4)_j\;
\mathbb{K}_6(p_1,p_2,p_3,p_4;h,\theta)}\\[8pt]
{\quad\quad -\frac{1}{4}\,\theta^{ij}\theta^{kl}\Big[
[2(p_{2 l}\,\delta_k^{\mu_1}\delta_i^{\mu_2}+p_{1 l}\,\delta_k^{\mu_2}\delta_i^{\mu_1})-(p_2\!-\!p_1)_{i}\,\delta_k^{\mu_1} \delta_l^{\mu_2}]\,\delta_{j}^{\mu_3}\,
\mathbb{K}_7(p_1,p_2,p_3,p_4;h,\theta)}\\[8pt]
{\phantom{\quad\quad -\frac{1}{4}\,\theta^{ij}\theta^{kl}\Big[}
+\delta_i^{\mu_1}\,
[2(p_{3 l}\,\delta_k^{\mu_2}\delta_j^{\mu_3}+p_{2 l}\,\delta_j^{\mu_2}\delta_k^{\mu_3})-(p_3\!-\!p_2)_{j}\,\delta_k^{\mu_2} \delta_l^{\mu_3}]
\;\mathbb{K}_8(p_1,p_2,p_3,p_4;h,\theta) }\\[8pt]
{\phantom{\quad\quad -\frac{1}{4}\,\theta^{ij}\theta^{kl}\Big[}
+\,2\,\delta_i^{\mu_1}\,\delta_j^{\mu_2}\,\delta_k^{\mu_3}\,p_{4l}\;\mathbb{K}_9(p_1,p_2,p_3,p_4;h,\theta)\Big],}
\end{array}
\end{equation*}

\begin{equation*}
\begin{array}{l}
{\mathbb{M}^{(2,1)}[(\mu_1,p_1);(\mu_2,p_2);(\mu_3,p_3);p_4;h\theta]=
\left[\frac{1}{2}\theta^{ij}\theta^{kl}(2p_{2l}\delta_{k}^{\mu_{1}}\delta_{i}^{\mu_{2}}+2p_{1l}\delta_{i}^{\mu_{1}}\delta_{k}^{\mu_{2}}-(p_{1}-p_{2})_{i}
\delta^{\mu_{1}}_{l}\delta_{k}^{\mu_{2}})\delta_{j}^{\mu_{3}}\right]}\\[8pt]
{\displaystyle{\frac{1}{p_{1}\wedge p_{2}}
\left(\frac{e^{-i\frac{h}{2}[p_1\wedge(p_2+p_3+p_4)+p_2\wedge(p_3+p_4)+p_4\wedge p_3]}-1}
{p_1\wedge(p_2+p_3+p_4)+p_2\wedge(p_3+p_4)+p_4\wedge p_3}-\frac{e^{-i\frac{h}{2}[p_1\wedge(p_3+p_4)+p_2\wedge(p_3+p_4)+p_4\wedge p_3]}-1}{p_1\wedge(p_3+p_4)+p_2\wedge(p_3+p_4)+p_4\wedge p_3}\right)}}\\[8pt]
{+\theta^{ij}\theta^{kl}\delta_{i}^{\mu_{1}}\delta_{k}^{\mu_{2}}\delta_{j}^{\mu_{3}}p_{4l}}\\[8pt]
{\displaystyle{\frac{1}{p_{2}\wedge p_{4}}
\left(\frac{e^{-i\frac{h}{2}[p_1\wedge(p_2+p_3+p_4)+p_2\wedge(p_3+p_4)+p_4\wedge p_3]}-1}
{p_1\wedge(p_2+p_3+p_4)+p_2\wedge(p_3+p_4)+p_4\wedge p_3}-\frac{e^{-i\frac{h}{2}[p_1\wedge(p_2+p_3+p_4)+(p_2+p_4)\wedge p_3]}-1}{p_1\wedge(p_2+p_3+p_4)+(p_2+p_4)\wedge p_3}\right)}}\\[8pt]
{\displaystyle
+\frac{1}{2}\theta^{ij}\theta^{kl}(2p_{2l}\delta_{k}^{\mu_{1}}\delta_{i}^{\mu_{2}}-p_{2i}\delta_{k}^{\mu_{1}}\delta_{l}^{\mu_{2}}+2p_{1l}\delta_{i}^{\mu_{1}}\delta_{k}^{\mu_{2}}-p_{1i}\delta_{l}^{\mu_{1}}\delta_{k}^{\mu_{2}})(p_{3}+p_{4})_{j}\theta^{mn}p_{4n}\delta^{\mu_{3}}_{m}\frac{1}{(p_{1}\wedge p_{2})(p_{4}\wedge p_{3})}}\\[8pt]
{\displaystyle{\Bigg(
\frac{e^{-i\frac{h}{2}[p_{1}\wedge p_{2}+p_{4}\wedge p_{3}+(p_{1}+p_{2})\wedge(p_{3}+p_{4})]}-1}{p_{1}\wedge p_{2}+p_{4}\wedge p_{3}+(p_{1}+p_{2})\wedge(p_{3}+p_{4})}-\frac{e^{-i\frac{h}{2}[p_{1}\wedge p_{2}+(p_{1}+p_{2})\wedge(p_{3}+p_{4})]}-1}{p_{1}\wedge p_{2}+(p_{1}+p_{2})\wedge(p_{3}+p_{4})}}}\\[8pt]
{\phantom{\frac{e^{-i\frac{h}{2}[p_{1}\wedge p_{2}+p_{4}\wedge p_{3}+(p_{1}+p_{2})\wedge(p_{3}+p_{4})]}-1}{p_{1}\wedge p_{2}+p_{4}\wedge p_{3}+(p_{1}+p_{2})\wedge(p_{3}+p_{4})}}\quad\quad\quad\quad
\displaystyle{ -\frac{e^{-i\frac{h}{2}[p_{4}\wedge p_{3}+(p_{1}+p_{2})\wedge(p_{3}+p_{4})]}-1}{p_{4}\wedge p_{3}+(p_{1}+p_{2})\wedge(p_{3}+p_{4})}
+\frac{e^{-i\frac{h}{2}(p_{1}+p_{2})\wedge(p_{3}+p_{4})}-1}{(p_{1}+p_{2})\wedge(p_{3}+p_{4})}
}\Bigg)
}\\[8pt]
{\displaystyle+\theta^{ij}\delta_{i}^{\mu_{1}}(p_{2}+p_{3}+p_{4})_{j}}\\[8pt]
{\displaystyle{\Bigg[\Bigg(\theta^{kl}\theta^{mn}\delta_{m}^{\mu_{2}}\delta_{k}^{\mu_{3}}(p_{2}+p_{4})_{l}\ p_{4n}\frac{1}{p_{2}\wedge p_{4}}-\theta^{kl}\theta^{mn}(p_{3}+p_{4})_{l}\ p_{4n}\delta_{k}^{\mu_{2}}\delta_{m}^{\mu_{3}}\frac{1}{p_{3}\wedge p_{4}}+\theta^{kl}\delta_{k}^{\mu_{2}}\delta_{l}^{\mu_{3}}\Bigg)}}\\[8pt]
{\displaystyle{\frac{1}{p_{2}\wedge( p_{3}+ p_{4})+p_{4}\wedge p_{3}}\Bigg(\frac{e^{-i\frac{h}{2}[p_{2}\wedge( p_{3}+ p_{4})+p_{4}\wedge p_{3}+p_{1}\wedge(p_{2}+p_{3}+p_{4})]}-1}{p_{2}\wedge( p_{3}+p_{4})+p_{4}\wedge p_{3}+p_{1}\wedge(p_{2}+p_{3}+p_{4})}-\frac{e^{-i\frac{h}{2}[p_{1}\wedge(p_{2}+p_{3}+p_{4})]}-1}{p_{1}\wedge(p_{2}+p_{3}+p_{4})}\Bigg)}}\\[8pt]
{\displaystyle{-\theta^{kl}\theta^{mn}\delta_{m}^{\mu_{2}}\delta_{k}^{\mu_{3}}(p_{2}+p_{4})_{l}p_{4n}}}\\[8pt]
{\quad\quad\quad\displaystyle{\frac{1}{p_{2}\wedge p_{4}}\frac{1}{(p_{2}+p_{4})\wedge p_{3}}\Bigg(\frac{e^{-i\frac{h}{2}[(p_{2}+p_{4})\wedge p_{3}+p_{1}\wedge(p_{2}+p_{3}+p_{4})]}-1}{(p_{2}+p_{4})\wedge p_{3}+p_{1}\wedge(p_{2}+p_{3}+p_{4})}-\frac{e^{-i\frac{h}{2}[p_{1}\wedge(p_{2}+p_{3}+p_{4})]}-1}{p_{1}\wedge(p_{2}+p_{3}+p_{4})}\Bigg)}}\\[8pt]
{\displaystyle{+\theta^{kl}\theta^{mn}(p_{3}+p_{4})_{l}\ p_{4n}\delta_{k}^{\mu_{2}}\delta_{m}^{\mu_{3}}}}\\[8pt]
{\quad\quad\quad\displaystyle{\frac{1}{p_{3}\wedge p_{4}}\frac{1}{p_{2}\wedge (p_{3}+ p_{4})}\Bigg(\frac{e^{-i\frac{h}{2}[p_{2}\wedge(p_{3}+p_{4})+p_{1}\wedge(p_{2}+p_{3}+p_{4})]}-1}{p_{2}\wedge(p_{3}+p_{4})+p_{1}\wedge(p_{2}+p_{3}+p_{4})}-\frac{e^{-i\frac{h}{2}[p_{1}\wedge(p_{2}+p_{3}+p_{4})]}-1}{p_{1}\wedge(p_{2}+p_{3}+p_{4})}\Bigg)\Bigg]}}\\[8pt]

{-\frac{1}{2}\theta^{ij}\theta^{kl}\delta_{i}^{\mu_{1}}\delta_{j}^{\mu_{2}}\delta_{k}^{\mu_{3}}p_{4l}}\\[8pt]
{\displaystyle{\frac{1}{p_{4}\wedge p_{3}}
\left(\frac{e^{-i\frac{h}{2}[p_1\wedge(p_2+p_3+p_4)+p_2\wedge(p_3+p_4)+p_4\wedge p_3]}-1}
{p_1\wedge(p_2+p_3+p_4)+p_2\wedge(p_3+p_4)+p_4\wedge p_3}-\frac{e^{-i\frac{h}{2}[p_1\wedge(p_2+p_3+p_4)+p_2\wedge(p_3+p_4)]}-1}{p_1\wedge(p_2+p_3+p_4)+p_2\wedge(p_3+p_4)}\right),}}\\[8pt]
\end{array}
\end{equation*}
\begin{equation*}
\mathbb{M}^{(1,2)}[(\mu_1,p_1);(\mu_2,p_2);(\mu_3,p_3);p_4;h\theta]=\overline{\mathbb{M}^{(2,1)}}[(\mu_3,-p_3);(\mu_2,-p_2);(\mu_1,-p_1);-p_4;h\theta],
\end{equation*}
\begin{equation*}
\mathbb{M}^{(0,3)}[(\mu_1,p_1);(\mu_2,p_2);(\mu_3,p_3);p_4;h\theta]=\overline{\mathbb{M}^{(3,0)}}[(\mu_3,-p_3);(\mu_2,-p_2);(\mu_1,-p_1);-p_4;h\theta].
\end{equation*}
The bar above $\mathbb{M}^{(1,2)}$ and $\mathbb{M}^{(3,0)}$  denotes complex conjugation.

\section{Acknowledgements}
This work has been financially supported in part by MICINN through grant
FPA2011-24560 and MPNS COST Action MP1405

%%%%%%%%%%%%%%%%%%%%%%%%%%%%%%%%%%%%%%%%%%%%%%%%%%%%%%%%
%%%%%%%%%%%%%%%%%%%%%%%%%%%%%%%%%%%%%%%%%%%%%%%%%%%%%%%%%
%%%%%%%%%%%%%%%%%%%%%%%%%%%%%%%%%%%%%%%%%%%%%%%%%%

\end{document}